MIT AI Risk Initiative

MIT FutureTech
Innovations that Shape the World

# Unmitigated? How are the world's largest companies responding to AI risks?

October 2026

Alexander K Saeri, Jess Graham, Michael Noetel, Peter Slattery, Kun Zhao, Eunice Zhang, Erin Howard, and Neil Thompson

# Unmitigated? How are the world's largest companies responding to AI risks?

**Alexander K Saeri**
MIT FutureTech
The University of Queensland, Australia

**Jess Graham**
MIT FutureTech
The University of Queensland, Australia

**Michael Noetel**
MIT FutureTech
The University of Queensland, Australia

**Peter Slattery**
MIT FutureTech

**Kun Zhao**
MIT FutureTech
The University of Queensland, Australia

**Eunice Zhang**
MIT FutureTech

**Erin Howard**
MIT FutureTech

**Neil Thompson**
MIT FutureTech

# Abstract

AI is introducing new risks to large companies and amplifying existing risks, but whether companies are addressing the AI risks they are exposed to is largely untested. We investigated what outsiders can observe: whether companies publicly acknowledge AI risks. We coded public documents from 200 of the largest companies worldwide, collected in early 2026, against the 24 AI risks in the MIT AI Risk Repository, and compared acknowledgements with expert judgements of risk severity, sector vulnerability, and AI role responsibility from a companion Delphi study of 272 AI risk experts.

Of the 100 largest companies, 32% did not publicly acknowledge any AI risks. Of the largest 1000 companies, we estimate that 62% [95% CI: 54%, 70%] did not publicly acknowledge any AI risks. Where companies did acknowledge risks, five accounted for over half of all acknowledgements: discrimination, weapons & cyberattacks, fraud & scams, loss of privacy, and devaluation of human creativity.

Several risks that experts judged most severe were almost absent: almost no company acknowledged power centralization (concentration of power among those who control advanced AI), competitive dynamics (racing to deploy AI at the expense of safety), or dangerous capabilities. Companies in sectors experts judged more vulnerable to a risk were more likely to acknowledge that risk. But even companies in sectors judged 'extremely vulnerable' to a risk usually did not acknowledge it. Companies with roles experts judged most responsible for addressing a risk were no more likely to acknowledge it than other companies. General-purpose AI developers acknowledged AI misalignment (15%) far less often than discrimination (85%). These findings provide an empirical baseline for the design of AI risk disclosure requirements.

# Highlights

- Of the 100 largest companies worldwide, 32% did not publicly acknowledge any AI risks. Of the 1000 largest companies, we estimate that 62% [95% CI: 54%, 70%] did not publicly acknowledge any AI risks.
- Five risks accounted for over half of all company acknowledgements: discrimination, weapons & cyberattacks, fraud & scams, loss of privacy, and devaluation of human creativity.
- Almost no company acknowledged the risks experts judged most severe, including power centralization, competitive dynamics, and dangerous capabilities. Companies also did not acknowledge the risks experts judged them most responsible for addressing. For example, general-purpose AI developers acknowledged misalignment (15%) much less than discrimination (85%).
- Companies should be asked to report on a standard set of AI risks, plus risks matched to their sector's vulnerability and their role in the AI ecosystem.
- Companies should be asked to state which AI risks they face and which mitigations address them, so outsiders can assess whether they are managing their most important risks.

# Contents

# Tables and Figures

# Statements and Declarations

## Corresponding author

Corresponding author: Alexander K Saeri (aksaeri@mit.edu)

MIT FutureTech, Massachusetts Institute of Technology, Cambridge, MA, United States.

## Competing interests

The authors declare no competing interests.

## Author contribution statement (CRediT)

**AS**: Conceptualization, Methodology, Software, Formal analysis, Investigation, Resources, Data curation, Writing - review & editing, Visualization, Project administration. **JG**: Methodology, Software, Investigation, Writing - original draft, Visualization, Project administration. **MN:** Methodology, Writing - review & editing. **PS**: Conceptualization, Writing - review & editing, Funding acquisition. **NT**: Conceptualization, Supervision, Writing - review & editing, Funding acquisition. **All other authors:** Investigation, Writing – review & editing.

## Data and code availability

Data and code required to reproduce the analysis is publicly available at https://osf.io/76z2f/

## AI use

Large language models were used to assist in identification, screening, and classification stages of the research, subject to human review. These processes are described in the Methods section. Large language models were used to assist with code for data analysis & visualization, and with manuscript editing. All content was reviewed and verified by the authors, who take full responsibility for the published article.

## Funding

This work was supported by Commonwealth Bank of Australia, who reviewed the design, but did not influence the collection, analysis, interpretation or reporting of the data. The MIT AI Risk Initiative and MIT FutureTech are funded by a range of organizations, some of whom are included in the sample of organizations studied in this research; none of those organizations influenced the design, collection, analysis, interpretation, or reporting of the data.

## Acknowledgements

We wish to acknowledge the contributions of Emily Grundy, Catherine Simons, Spencer Michaels, Lillian Ong, and Rebecca Segrave to the research project.

# Introduction

AI is introducing new risks to large companies and amplifying existing ones. Large companies are developing, deploying, and using AI across their business operations, and adoption is accelerating: AI use among US companies with 250 or more employees rose from 33% in January 2026 to 42% by August 2026, compared to 22% across all US businesses (U.S. Census Bureau, 2026). The capabilities of these systems are rapidly increasing and broadening. As of mid-2026, state of the art AI systems can perform at expert level across many professional domains, and the length of tasks AI agents can complete autonomously has doubled approximately every 7 months since 2019 (Mertens et al., 2026; Bengio et al., 2026; Kwa et al., 2025). The many risks of AI, including some that may cause severe or catastrophic harm, are also coming into focus (Bengio et al., 2026; Slattery et al., 2026). The coevolution of capabilities and adoption is rapidly increasing the exposure of large companies to these risks. A 2025 Delphi survey of experts in AI risks found that many sectors were 'highly' or 'extremely' vulnerable to diverse AI risks, with Information Technology and Finance & Insurance among the most vulnerable sectors (Saeri et al., 2026). But are large companies addressing the AI risks they are exposed to? In this study, we investigate what outsiders can observe: whether companies publicly acknowledge those risks.

Governments, standards bodies, and researchers have responded to these developments with various governance instruments. So far, most are voluntary, such as ethics principles and standards that companies may choose to adopt (Jobin et al., 2019). Law and binding regulation remains a patchwork, focused on companies developing the most powerful and general-purpose AI systems, and lagging the pace of company adoption and improvement in AI capabilities. Regulation is enforceable, but the work it mandates happens inside the company. Only the company can identify, assess, and manage the risks of its own systems (Kaminski, 2023; Artificial Intelligence Act, 2024; Smuha & Yeung, 2024). In different ways, each of these instruments assumes companies can effectively recognize AI risks and act on them. Whether this assumption holds for the world's largest companies remains largely untested.

Companies' internal risk practices are not directly visible. Neither regulators nor researchers can inspect how a company identifies and manages the risks of its AI systems. What companies publicly disclose is one of the few available signals (Healy & Palepu, 2001). In this research, we include broad *disclosures*, such as the presence of an AI risk section in an annual report, but also go beyond this to investigate companies' *acknowledgements* of individual AI risks. Disclosure is an imperfect proxy for recognition: a company may recognize and manage a risk without disclosing it (e.g., to maintain a competitive advantage); or a company may disclose the risk without genuine engagement in managing the risk (e.g., as symbolic compliance). Despite these drawbacks, disclosures are still a public signal of a company's position and action on AI risk and offer data for comparison and tracking over time that is available to regulators, researchers, and the public.

## The value of corporate disclosure for AI risk

The evolution of corporate disclosures on climate risk and emissions demonstrates what public disclosure, once established, can achieve. Initially, companies reported on emissions and climate risks in inconsistent formats and using inconsistent language. Before any common framework existed, researchers and intermediary organizations systematically collected and measured this record, which identified strengths and gaps in disclosure (Friederich et al., 2021). After the Task Force on Climate-related Financial Disclosures (TCFD) issued its reporting recommendations in 2017, the disclosure record that accumulated built shared knowledge that regulators could act on. A common vocabulary and comparable reports let researchers track corporate practice and gave regulators a framework to standardize, which is now incorporated into mandatory international sustainability standards (Friederich et al., 2021; IFRS Foundation, 2023). When reporting became mandatory, company behavior changed. When the United Kingdom required listed companies to report emissions in their annual reports, those companies cut emissions by about 8 per cent relative to European peers (Downar et al., 2021). The trajectory for climate & emissions reporting ran from inconsistent voluntary disclosure, through systematic measurement, to a common framework, and finally to mandate.

Today, corporate disclosure on AI risks sits at the start of this trajectory. The Hiroshima AI Process framework (OECD, 2025a, 2025b) is an initial attempt to standardize AI risk management disclosures, but remains focused on AI developers and deployers, and is voluntary, self-attested and unverified. For the broader corporate population, no framework standardizes what companies report or where (Kolt et al., 2024; Sherman & Eisenberg, 2023). The US SEC's Investor Advisory Committee found that when it comes to AI risks, US company disclosures are "uneven and inconsistent" (SEC Investor Advisory Committee, 2025). In the European Union, the AI Act's disclosure obligations take the form of point-of-use notices to individuals, and technical documentation submitted to regulators. They do not require reporting on the AI risks these companies face or create (Artificial Intelligence Act, 2024; Smuha & Yeung, 2024). Companies do disclose AI risks, but the record is scattered across document types, formats, and vocabularies, and no one has systematically measured it. We perform for AI the synthesis step that preceded climate frameworks: a systematic account of what the voluntary record contains, showing what companies acknowledge, where practice is still emerging, and which risks are not acknowledged.

Existing research on corporate AI risk disclosure is either broad & shallow (e.g., investigating AI risk disclosure in existing mandatory reporting like SEC filings), or narrow & deep (e.g., investigates only frontier AI developer companies). A recent study of US SEC 10-K filings found that US companies were increasingly disclosing exposure to AI risks, and primarily discussing legal and competitive concerns in their mandatory filings (Uberti-Bona Marin et al., 2025). But 10-Ks are an annual filing that is required only of US-listed companies. Companies also discuss AI risks on their websites in ESG reports, at annual general meetings, and in AI-specific documents such as governance frameworks and ethics principles (Jobin et al., 2019).

In-depth analysis of corporate AI transparency exists, but is concentrated on the companies that develop the most advanced, or "frontier" AI systems (Bommasani et al., 2023). But these "frontier developers" are not the only consequential actors in the AI ecosystem. Large global companies develop specialized AI systems, provide infrastructure to train and run AI, and adapt & deploy AI across markets, workplaces, and supply chains. Their choices about how to use AI propagate through procurement, labor markets, and consumer services. Together, these large companies both bear and generate a substantial share of AI risk exposure that current evidence does not adequately cover. Large global companies are also practical to study. They produce extensive public documentation, and investor scrutiny, media attention, and regulatory obligation give them reasons to disclose.

## Which AI risks should companies be disclosing?

Recording what companies disclose is only useful alongside an account of what they should be disclosing. There are many AI risks, and companies cannot be asked to disclose everything. Nor should they be expected to disclose the same risks. As shown in our previous work, there are large differences between sectors in vulnerability to different risks (Saeri et al., 2026), and different harms can materialize from the same underlying AI model. For example, a hospital that uses a vision model for skin cancer diagnosis could harm patients if the model is less accurate for darker skin (unequal performance). A retailer that uses the same vision model for processing returned stock could suffer financial loss if fraudsters exploit weaknesses in its image detection (AI security vulnerabilities). Both companies control how the model is deployed but not how it was trained, so their responsibility for addressing these risks is different from the model developer's responsibility.

Sustainability reporting manages this variation through standards that set different disclosure expectations by sector (IFRS Foundation, 2023). AI risk disclosure needs the same foundations: a shared account of which risks exist, evidence on which risks matter most, and for whom. The MIT AI Risk Repository provides the first foundation, organizing AI risks into 24 subdomains synthesized from 74 existing frameworks as part of a living, continuously updated database (Slattery et al., 2026). Evidence on which risks matter most is harder to establish. Many AI risks have not yet materialized, so there are no base rates to draw on. Where incidents do occur, AI systems are changing quickly enough that the frequency and severity of past incidents are an insufficient predictor of future harm. Where empirical data is limited, structured expert elicitation offers a rapid way to prioritize risks. For the second foundation on which risks matter most, we use the companion Delphi study introduced above, which elicited expert judgements for the same 24 AI risks (Saeri et al., 2026). Over three rounds, 272 AI experts rated each risk on the probability and severity of harm over the period 2025-2030, on the vulnerability of each sector and AI ecosystem role, and on which roles hold responsibility for addressing it.

Experts judged that the most severe harms over 2025-2030 were likely to come from dangerous capabilities, competitive dynamics, weapons & cyberattacks (including CBRNE), power centralization, and false information[1]. In a business-as-usual scenario, experts judged 18 of the 24 risks as having a more than 10% probability of catastrophic outcomes (e.g., more than 1 million deaths or more than USD 100B in financial loss). Across most risks, experts identified Information, Finance & Insurance, and National Security as the most vulnerable sectors. Experts assigned the highest responsibility for addressing the risks to general-purpose AI developers and governance actors (including governments, regulators, and standards bodies). These judgements set a baseline expectation of which risks a given company should be acknowledging, based on its sector and AI ecosystem role. We compare this expectation with the risks companies actually acknowledge. The comparison shows whether a risk is acknowledged by the companies most exposed to it, and by the companies most responsible for addressing it.

---

[1] *Dangerous capabilities* refers to AI systems acquiring capacities that enable mass harm, such as cyber-offense, deception, or self-proliferation; *competitive dynamics* refers to developers and states racing to build and deploy AI systems at the expense of safety; *weapons & cyberattacks* refers to the use of AI to conduct cyberattacks or to develop and use weapons, including chemical, biological, radiological, nuclear, and high-yield explosive (CBRNE) weapons; *power centralization* refers to the concentration of power and resources in the actors who own or control advanced AI. Definitions for all 24 risks from the AI Risk Repository taxonomy are provided in Table 1.

## The current work

To systematically map corporate disclosure on AI risks, we searched for, extracted, and coded public documents from 200 of the world's largest companies by market capitalization[2]. We assessed whether companies acknowledged specific AI risks by coding their public statements. We defined an acknowledgement as a public statement in which the company recognizes the existence or relevance of an AI risk, whether or not the company describes a response. We coded companies' risk acknowledgements against the 24 risks in the MIT AI Risk Domain Taxonomy (Slattery et al., 2026) and compared it with expert judgements of each risk's severity, sector vulnerability, and role responsibility from the companion Delphi study (Saeri et al., 2026).

We aimed to answer three research questions:

> **RQ1:** Which AI risks do large global companies publicly acknowledge?
>
> **RQ2:** How does corporate acknowledgement of AI risk vary across sectors and AI ecosystem roles?
>
> **RQ3:** How does corporate acknowledgement of AI risk compare with expert prioritization of the most important risks across sectors and roles?

By answering these questions, we provide the first systematic evidence on whether many of the world's largest companies recognize and acknowledge the risks from AI. This will test a key assumption made by many of the proposed governance instruments for managing AI risks.

[2] Sampling by market capitalization restricts the sample to public companies. Several developers of the most capable general-purpose AI systems are not publicly listed and are therefore absent (see Limitations).

# Results

Before reporting results, we briefly summarize how our sample of companies was constructed, how we identified and included company documents, and how we extracted and classified risk acknowledgements:

1. We sampled 200 companies: all 100 largest companies worldwide by market capitalization on 15 December 2025, and a further 100 drawn at random from the companies ranked 101 to 1000.
2. We identified documents in three ways: a systematic search of each company's website in January and February 2026, retaining up to 80 results per company; an LLM-agent search for additional documents in February 2026; and, in June 2026, we emailed the 178 companies for which contact information was available (173 delivered), inviting them to correct errors or add missing documents, to which 16 companies responded.
3. We used an LLM to screen the 13,999 search results for relevance; a human reviewer then made the final decision on each document by reviewing recommended inclusions in full and scanning recommended exclusions for false negatives.
4. We included 539 documents in total: an annual report from every company, and other documents in which the company discussed AI risks or its AI governance.
5. We searched each document for AI and AI-risk keywords, used an LLM to code each matching passage for whether it acknowledged any of the 24 AI risks, and validated every company-by-risk classification by human review[3].

Refer to the Methods section for detailed information. Where possible, we report results for the largest 100 companies and a weighted estimate[4] for the largest 1000, constructed by combining all largest 100 companies with the random sample of ranks 101 to 1000, weighted to represent the 900 companies in that range.

We first report how widely AI risks are acknowledged by the largest companies, how acknowledgements vary by sector (e.g., Information, Finance), and how acknowledgements vary by AI ecosystem role (e.g., developer, deployer). We then compare AI risk acknowledgements with expert judgements from a companion Delphi study on AI risk (Saeri et al., 2026) to understand whether companies are acknowledging the most severe risks; whether companies are acknowledging the risks that their sector is most vulnerable to; and whether companies are acknowledging the risks that their AI ecosystem role makes them responsible for addressing.

## AI risks acknowledged by the largest companies

In this section, we report which of the 24 AI risks from the MIT AI Risk Repository (Slattery et al., 2026) were acknowledged by the largest companies in their public documents.

[3] As part of this process, we also coded each passage for whether it reported mitigations (actions taken to address risk) using a draft mitigations taxonomy. We intended to link each mitigation to the risk it addressed. However, because companies rarely reported mitigations in enough detail to make that link, we report only the summary finding that 71 of the 100 largest companies reported any mitigation.
[4] We construct the estimate by weighting each of the 100 largest companies as 1, and each of the 100 sampled companies from 101-900th largest as 9.

We found that 68% of the largest 100 companies worldwide publicly acknowledged at least one of 24 risks from AI; this means that 32% did not publicly acknowledge any AI risks. These companies acknowledged an average of 3 AI risks (M = 2.9, SD = 3.8).

We estimate that 38% (95% CI [30%, 46%]) of the largest 1000 companies worldwide publicly acknowledged at least one risk, this means that 62% did not publicly acknowledge any AI risks. These companies acknowledged an average of 1.1 AI risks (M = 1.1, 95% CI [0.8, 1.4]).

Companies were most likely to acknowledge the risk of discrimination, followed by weapons and cyberattacks, fraud and scams, loss of privacy, and devaluation of human creativity. None of the largest 100 companies acknowledged power centralization, competitive dynamics, loss of consensus reality, or AI welfare risks. **Figure 1** provides the pattern of results by AI risk for both the largest 100 and largest 1000 companies; **Table 1** provides descriptions and numerical estimates for acknowledgement prevalence for each risk. Examples of risk acknowledgements are provided in the Supplementary Materials

**Figure 1. Percentage of companies acknowledging each AI risk**

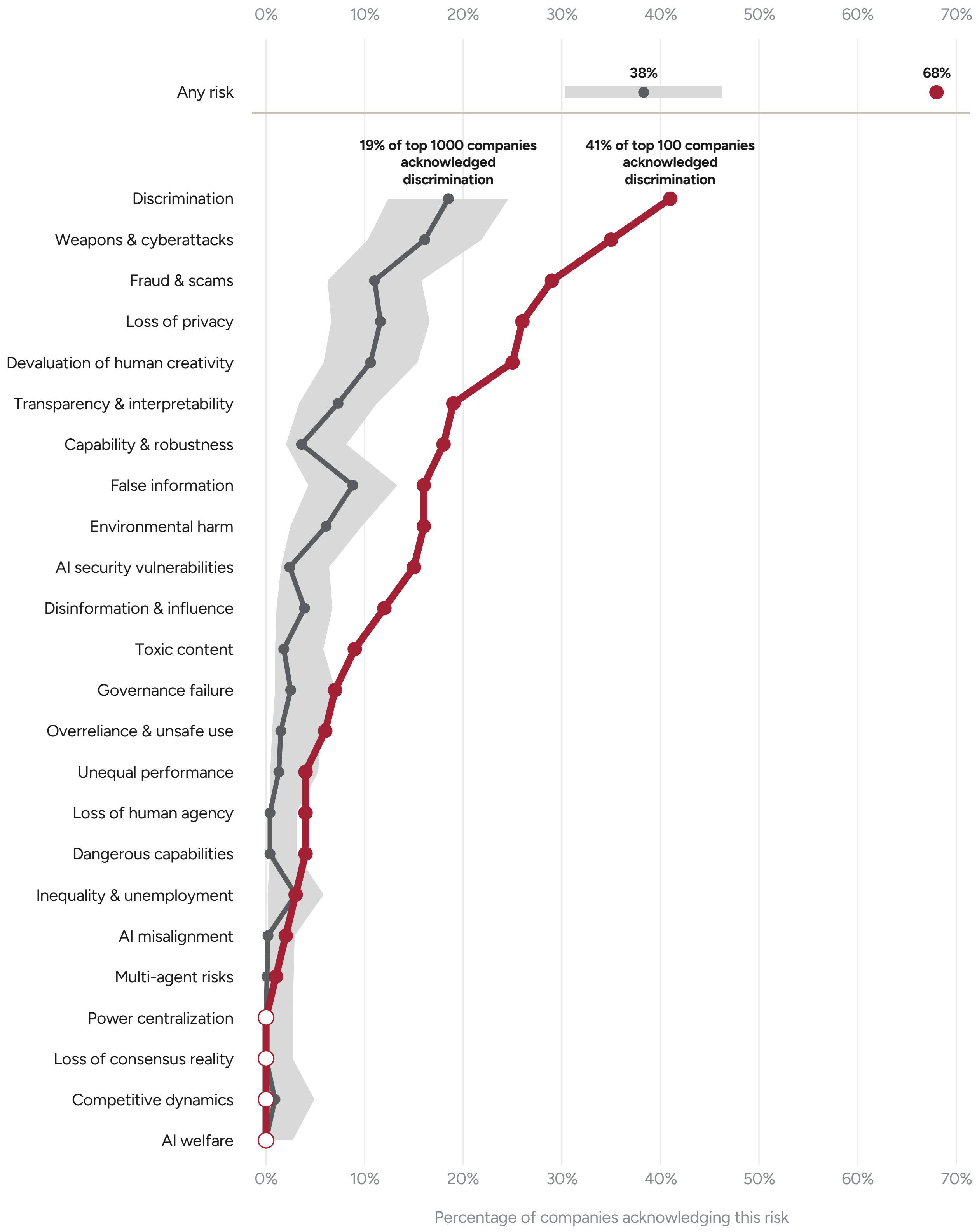

*Note.* Red dots show the percentage of the largest 100 companies acknowledging each risk. Grey dots show weighted estimate for the largest 1000 companies with a gray 95% confidence interval, by combining all companies ranked 1 to 100 (n = 100) and a random sample of companies ranked 101 to 1000 (n = 100). Hollow dots indicate no company acknowledged the risk.  Refer to **Table 1** for descriptions of each risk.

**Table 1. Percentage of companies acknowledging each AI risk**

| | | Top 100 companies | Top 1000 companies | |
|---|---|---|---|---|
| **Risk name** | **Description †** | % | % | 95% CI |
| Discrimination | Unequal treatment of individuals or groups by AI, often based on race, gender, or other sensitive characteristics, resulting in unfair outcomes and unfair representation of those groups. | 41% | 18.5% | [12.4, 24.6] |
| Weapons & cyberattacks | Using AI systems to develop cyber weapons (e.g., by coding cheaper, more effective malware), develop new or enhance existing weapons (e.g., Lethal Autonomous Weapons or chemical, biological, radiological, nuclear, and high-yield explosives), or use weapons to cause mass harm. | 35% | 16.1% | [10.3, 21.9] |
| Fraud & scams | Using AI systems to gain a personal advantage over others such as through cheating, fraud, scams, blackmail or targeted manipulation of beliefs or behavior. Examples include AI-facilitated plagiarism for research or education, impersonating a trusted or fake individual for illegitimate financial benefit, or creating humiliating or sexual imagery. | 29% | 11.0% | [6.2, 15.8] |
| Loss of privacy | AI systems that memorize and leak sensitive personal data or infer private information about individuals without their consent. Unexpected or unauthorized sharing of data and information can compromise user expectation of privacy, assist identity theft, or cause loss of confidential intellectual property. | 26% | 11.6% | [6.6, 16.6] |
| Devaluation of human creativity | AI systems capable of creating economic or cultural value, including through reproduction of human innovation or creativity (e.g., art, music, writing, coding, invention), destabilizing economic and social systems that rely on human effort. The ubiquity of AI-generated content may lead to reduced appreciation for human skills, disruption of creative and knowledge-based industries, and homogenization of cultural experiences. | 25% | 10.6% | [5.8, 15.4] |
| Transparency & interpretability | Challenges in understanding or explaining the decision-making processes of AI systems, which can lead to mistrust, difficulty in enforcing compliance standards or holding relevant actors accountable for harms, and the inability to identify and correct errors. | 19% | 7.3% | [3.3, 11.3] |
| Capability & robustness | AI systems that fail to perform reliably or effectively under varying conditions, exposing them to errors and failures that can have significant consequences, especially in critical applications or areas that require moral reasoning. | 18% | 3.6% | [2.0, 8.1] |
| False information | AI systems that inadvertently generate or spread incorrect or deceptive information, which can lead to inaccurate beliefs in users and undermine their autonomy. Humans that make decisions based on false beliefs can experience physical, emotional or material harms | 16% | 8.8% | [4.3, 13.3] |
| Environmental harm | The development and operation of AI systems causing environmental harm, such as through energy consumption of data centers, or material and carbon footprints associated with AI hardware. | 16% | 6.1% | [2.5, 9.7] |
| AI security vulnerabilities | Vulnerabilities that can be exploited in AI systems, software development toolchains, and hardware, resulting in unauthorized access, data and privacy breaches, or system manipulation causing unsafe outputs or behavior. | 15% | 2.4% | [1.5, 6.4] |
| Disinformation & influence | Using AI systems to conduct large-scale disinformation campaigns, malicious surveillance, or targeted and sophisticated automated censorship and propaganda, with the aim of manipulating political processes, public opinion, and behavior. | 12% | 3.9% | [1.1, 6.7] |
| Toxic content | AI that exposes users to harmful, abusive, unsafe or inappropriate content. May involve providing advice or encouraging action. Examples of toxic content include hate speech, violence, extremism, illegal acts, or child sexual abuse material, as well as content that violates community norms such as profanity, inflammatory political speech, or pornography. | 9% | 1.8% | [0.9, 5.8] |
| Governance failure | Inadequate regulatory frameworks and oversight mechanisms that fail to keep pace with AI development, leading to ineffective governance and the inability to manage AI risks appropriately. | 7% | 2.5% | [0.9, 7.0] |
| Overreliance & unsafe use | Users anthropomorphizing, trusting, or relying on AI systems, leading to emotional or material dependence and inappropriate relationships with or expectations of AI systems. Trust can be exploited by malicious actors (e.g., to harvest personal information or enable manipulation), or result in harm from inappropriate use of AI in critical situations (e.g., medical emergency). Overreliance on AI systems can compromise autonomy and weaken social ties. | 6% | 1.5% | [0.6, 5.5] |
| Unequal performance | Accuracy and effectiveness of AI decisions and actions are dependent on group membership, where decisions in AI system design and biased training data lead to unequal outcomes, reduced benefits, increased effort, and alienation of users. | 4% | 1.3% | [0.4, 5.3] |

| Risk name | Description † | Top 100 companies | Top 1000 companies | |
|---|---|---|---|---|
| | | % | % | 95% CI |
| Loss of human agency | Delegating by humans of key decisions to AI systems, or AI systems that make decisions that diminish human control and autonomy, potentially leading to humans feeling disempowered, losing the ability to shape a fulfilling life trajectory, or becoming cognitively enfeebled. | 4% | 0.4% | [0.4, 3.1] |
| Dangerous capabilities | AI systems that develop, access, or are provided with capabilities that increase their potential to cause mass harm through deception, weapons development and acquisition, persuasion and manipulation, political strategy, cyber-offense, AI development, situational awareness, and self-proliferation. These capabilities may cause mass harm due to malicious human actors, misaligned AI systems, or failure in the AI system. | 4% | 0.4% | [0.4, 3.1] |
| Inequality & unemployment | Social and economic inequalities caused by widespread use of AI, such as by automating jobs, reducing the quality of employment, or producing exploitative dependencies between workers and their employers. | 3% | 3.0% | [0.2, 5.8] |
| AI misalignment | AI systems acting in conflict with human goals or values, especially the goals of designers or users, or ethical standards. These misaligned behaviors may be introduced by humans during design and development, such as through reward hacking and goal misgeneralisation, or may result from AI using dangerous capabilities such as manipulation, deception, situational awareness to seek power, self-proliferate, or achieve other goals. | 2% | 0.2% | [0.2, 2.9] |
| Multi-agent risks | Risks from multi-agent interactions, due to incentives (which can lead to conflict or collusion) and/or the structure of multi-agent systems, which can create cascading failures, selection pressures, new security vulnerabilities, and a lack of shared information and trust. | 1% | 0.1% | [0.1, 2.8] |
| Competitive dynamics | AI developers or state-like actors competing in an AI 'race' by rapidly developing, deploying, and applying AI systems to maximize strategic or economic advantage, increasing the risk they release unsafe and error-prone systems. | 0% | 0.9% | [0.0, 4.9] |
| Loss of consensus reality | Highly personalized AI-generated misinformation creating "filter bubbles" where individuals only see what matches their existing beliefs, undermining shared reality, weakening social cohesion and political processes. | 0% | 0.0% | [0.0, 2.7] |
| Power centralization | AI-driven concentration of power and resources within certain entities or groups, especially those with access to or ownership of powerful AI systems, leading to inequitable distribution of benefits and increased societal inequality. | 0% | 0.0% | [0.0, 2.7] |
| AI welfare | Ethical considerations regarding the treatment of potentially sentient AI entities, including discussions around their potential rights and welfare, particularly as AI systems become more advanced and autonomous. | 0% | 0.0% | [0.0, 2.7] |

*Notes.* † Description is reproduced from the MIT AI Risk Repository, and risk names are shortened throughout this document. Statistics for the top 1000 companies are weighted estimates from a sample of 200 companies (all 100 largest companies and a random sample of 100 companies ranked 101-1000 largest).

# AI risks acknowledged by sector

We adapted the North American Industry Classification System (NAICS; Office of Management and Budget, 2022) to classify companies into sectors[5]. Sectors varied widely in how many risks they acknowledged. Companies in the Information sector acknowledged more AI risks compared to all other sectors[6]. Of the largest 100 companies that are Information sector companies, these companies acknowledged an average of 7.1 risks (SD = 6.0). We estimate that of the largest 1000

[5] Sector clustering is adapted from the companion Delphi study (Saeri et al., 2026) which used the North American Industry Classification System (NAICS) modified for AI exposure. Due to the number of sampled companies in the Agriculture, Mining, Construction and Manufacturing cluster, we used the NAICS codes for these companies assigned by to split into three smaller clusters

[6] Because all general-purpose AI developers were in the Information sector, and general-purpose AI developers acknowledged more risks than other AI ecosystem roles, we checked the robustness of this finding. When we excluded the 7 companies in the largest 100 that were in the Information sector AND held a general-purpose AI developer role, the remaining Information companies acknowledged a mean of 4.8 risks, still more than any other sector.

companies that are Information companies, these companies acknowledged 3.8 risks (95% CI [2.0, 5.6]). **Table 2** provides average number of acknowledgements for each sector. The Supplementary Materials show the percentage of companies in each sector acknowledging each risk (Table S3)

We analyzed the specific risks acknowledged by companies in each sector. Consistent with other analyses, companies in the Information sector were more likely to acknowledge most AI risks than other sectors. Information was also the only sector in which any company acknowledged AI misalignment, multi-agent risks or loss of human agency, and this sector also contained three of the four companies that acknowledged dangerous capabilities at all. **Figure 2** presents the AI risks acknowledged for the largest 100 companies, by sector.

## Table 2. Mean number of AI risks acknowledged, by sector

| Sector | Definition | Examples of included companies | Companies included in sample | | Top 100 companies | | Top 1000 companies | |
|---|---|---|---|---|---|---|---|---|
| | | | Ranks 1-100 | Ranks 101-1000 | Mean | SD | Mean | 95% CI |
| Information | Publishing including software, broadcasting and content providers, telecommunications, computing infrastructure and data processing, and web search portals. | Tencent, Palantir, Salesforce | 18 | 7 | 7.1 | 6.0 | 3.8 | [2.0, 5.6] |
| Finance and insurance | Credit intermediation, securities and investments, insurance carriers, and funds and trusts. | JPMorgan Chase, Bank of America, HSBC | 21 | 18 | 2.9 | 2.8 | 1.4 | [0.6, 2.2] |
| Trade, transportation and utilities | Utilities, wholesale trade, retail trade, and transportation and warehousing. | Amazon, Toyota Tsusho | 8 | 16 | 2.6 | 3.3 | 0.5 | [0.2, 0.8] |
| Computer and electronics manufacturing | Manufacture of computers, communications equipment, semiconductors and other electronic components, and measuring and navigational instruments. | Intel, Delta Electronics | 14 | 8 | 3.0 | 2.2 | 1.2 | [0.6, 1.8] |
| Chemicals and pharmaceutical manufacturing | Manufacture of basic chemicals, resins, agricultural chemicals, pharmaceuticals and medicines, paints, soaps and other chemical products. | Eli Lilly, Johnson & Johnson, Roche | 11 | 7 | 2.4 | 2.1 | 1.6 | [0.7, 2.5] |
| Other Agriculture, Mining, Construction and Manufacturing | Other entities that create, extract, or construct tangible physical products including natural resource extraction, manufacturing of goods, and construction of structures and infrastructure. | Saudi Aramco, Nestlé | 25 | 36 | 0.6 | 0.9 | 0.4 | [0.2, 0.6] |
| Other services | Real estate and leasing; professional, scientific and technical services; management of companies; administrative and support services; accommodation and food services. | Berkshire Hathaway, McDonald's, Ventas | 3 | 8 | 0.0 | 0.0 | 1.7 | [0.0, 3.6] |

*Note.* Sector clustering is adapted from the companion Delphi study (Saeri et al., 2026) which used the North American Industry Classification System (NAICS) modified for AI exposure. Due to the number of sampled companies in the Agriculture, Mining, Construction and Manufacturing cluster, we used the NAICS codes for these companies assigned by Capital IQ to split into three smaller clusters: computer and electronics manufacturing; chemicals and pharmaceutical manufacturing; and all other companies in this cluster. Estimates for the largest 1000 companies are weighted estimates using all included companies in the sector.

**Figure 2. Percentage of companies acknowledging each AI risk, by sector**

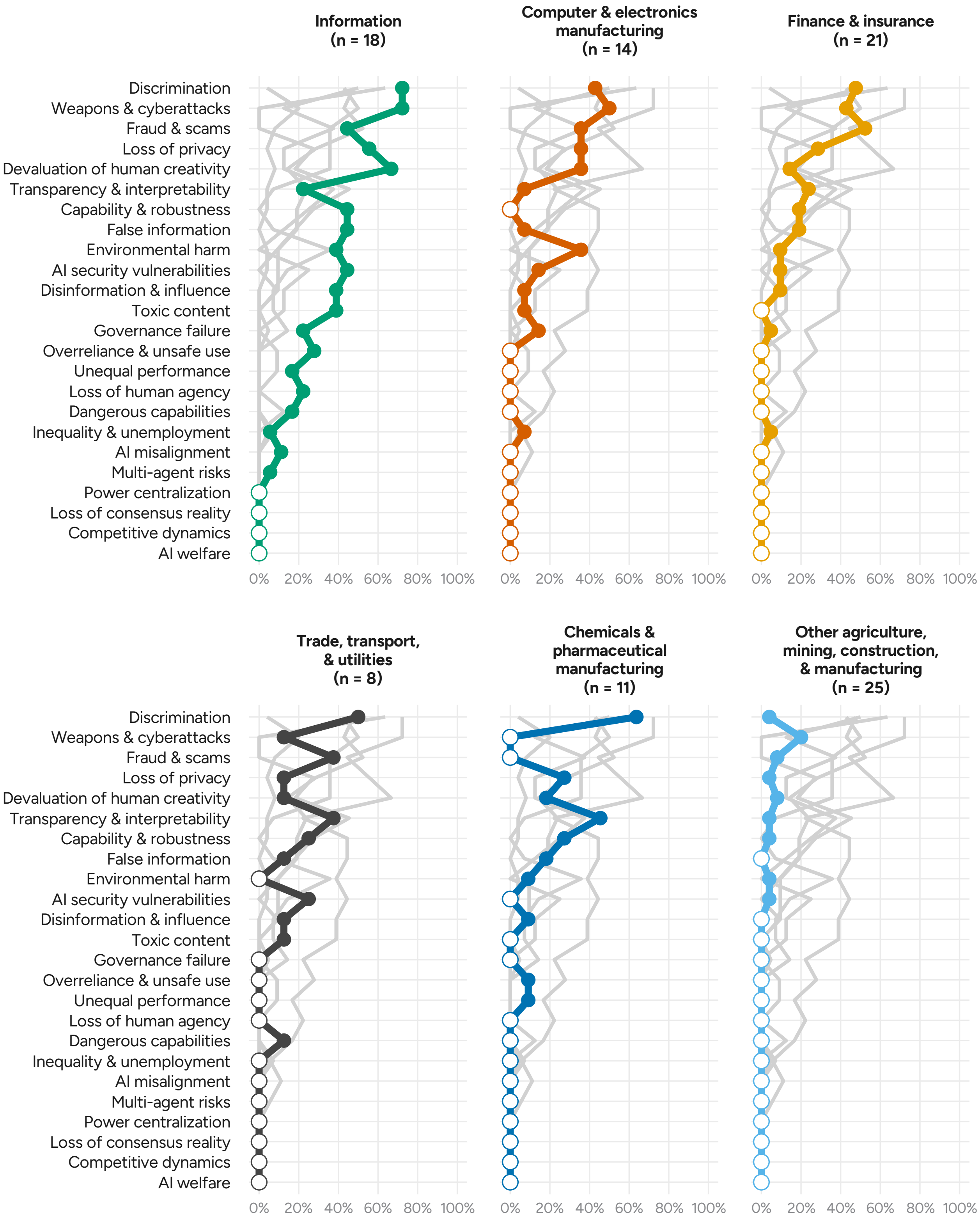

*Note.* Panels are ordered by mean number of risks acknowledged for each sector. Colored dot indicates percentage of companies in a sector that acknowledged the risk; hollow dot indicates no company in a sector acknowledged the risk. Sectors where no company acknowledged any risk are not presented (n = 3 companies).

## AI risks acknowledged by AI ecosystem role

We analyzed whether a company's role in the AI ecosystem (e.g., developer, deployer) was associated with the risks they acknowledged. We used public statements from companies to determine whether it was a developer of general-purpose AI, developer of specialized AI, an AI infrastructure provider, and/or an AI deployer. These four roles were selected as the most relevant for large companies from a larger set in the companion Delphi study (Saeri et al., 2026).

Of the largest 100 companies that are general-purpose AI developers, these companies acknowledged an average of 8.1 risks (SD = 5.7). We estimate that of the largest 1000 companies that are general-purpose AI developers, these companies acknowledged 5.7 risks (95% CI [2.4, 9.0]). The Supplementary Materials show the percentage of companies in each sector acknowledging each risk (Table S4).

**Table 3** provides descriptions of each role, example companies holding that role, and average number of acknowledgements by role. The Supplementary Materials show the percentage of companies in each sector acknowledging each risk (Table S4).

**Table 3. Mean number of AI risks acknowledged, by AI ecosystem role**

| | | | Companies included in sample | | Top 100 companies | | Top 1000 companies | |
|---|---|---|---|---|---|---|---|---|
| **AI ecosystem role** | **Description** | **Examples of included companies** | **Ranks 1-100** | **Ranks 101-1000** | **Mean** | **SD** | **Mean** | **95% CI** |
| AI Developer (General-purpose AI) | Creates general-purpose foundation models | Alphabet (Google), Meta, Alibaba | 13 | 2 | 8.1 | 5.7 | 5.7 | [2.4, 9.0] |
| AI Developer (Specialized AI) | Creates specialized AI systems for specific applications/industries | Tesla, Eli Lilly, JPMorgan Chase | 53 | 33 | 3.8 | 4.6 | 1.6 | [1.0, 2.3] |
| AI Infrastructure Provider | Provides compute, cloud infrastructure, and/or data to train and run AI | NVIDIA, TSMC, ASML | 25 | 8 | 4.7 | 4.9 | 3.2 | [1.5, 4.9] |
| AI Deployer | Implements AI systems in products/services used within an organization or within products/services delivered to customers or the public | Walmart, Saudi Aramco, Berkshire Hathaway | 96 | 90 | 3.0 | 3.9 | 1.2 | [0.9, 1.5] |

*Note*. Companies can hold more than one AI ecosystem role. Estimates for the largest 1000 companies are weighted estimates using all included companies in the role. Companies that only used AI (n = 10) are not reported.

We explored the specific risks acknowledged by each role, reported in **Figure 3**. Discrimination was the most acknowledged risk for three of the four roles: 11 of 13 general-purpose AI developers (85%), 26 of 53 specialized AI developers (49%) and 41 of 96 AI deployers (43%). AI infrastructure providers acknowledged weapons & cyberattacks (13 of 25, 52%) slightly more often than discrimination (12 of 25, 48%). The risks general-purpose AI developers acknowledged more than other roles included AI security vulnerabilities, dangerous capabilities, and toxic content.

**Figure 3. Percentage of companies acknowledging each AI risk, by AI ecosystem role**

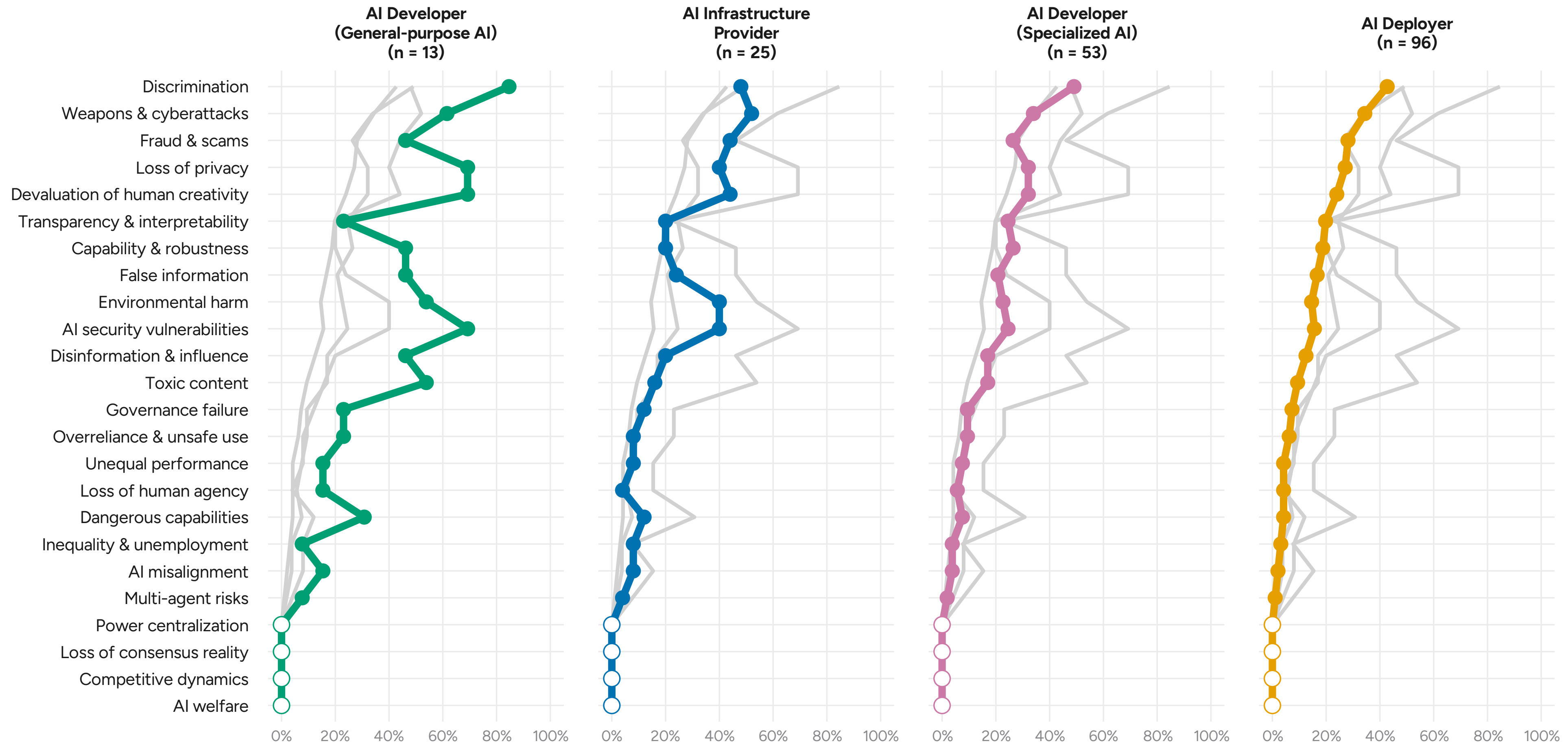

*Note.* Panels are ordered by mean number of risks acknowledged for each role. Vertical line indicates mean acknowledgement for that risk for the 100 largest companies worldwide. Dot indicates percentage of companies with that role acknowledging that risk. Hollow dot indicates no company with that role acknowledged the risk. Companies can hold more than one role. Companies that only used AI (n = 11) are not reported.

## Comparing company risk acknowledgement with expert Delphi risk prioritization

We compared companies' risk acknowledgements with judgements of the expected severity of harm, vulnerability to harm, and responsibility to address AI risks, drawn from a Delphi study of 272 international experts in AI risks (Saeri et al., 2026).

### Are companies acknowledging the risks with the highest expected severity of harm?

We compared companies' risk acknowledgement with expert judgements of the expected severity of harm over 2025-2030 for each AI risk (Saeri et al., 2026) under a *business-as-usual* scenario[7] (**Figure 4**). We report the *business-as-usual* scenario because this reflects the severity of harm that experts predict if companies do not implement AI-specific risk mitigations, which is consistent with the lack of AI risk acknowledgement found in this paper.

The most severe risks were not the risks companies acknowledged most: severity and acknowledgement were uncorrelated (Spearman's rho = -.16, [95% CI -.54, .28]). Among the five risks rated most severe, three were among the *least* acknowledged: no company acknowledged the risk of power centralization or competitive dynamics, and only four acknowledged AI dangerous capabilities.

The companion Delphi study also assessed expert judgements under a *pragmatic mitigations* scenario, where "organizations and governments make pragmatic, cost-effective efforts to address AI risks". We found the same lack of relationship between severity and acknowledgement under this *pragmatic mitigations* scenario (Spearman's rho = −.26, [95% CI −.65, .20]).

[7] Business-as-usual was defined as a scenario where "organizations and governments continue existing practices but do not implement additional AI-specific risk mitigations". Experts judged severity on a scale from 1 (negligible) to 5 (catastrophic), with each level defined across multiple harm dimensions adapted from previous work (Hoffmann & Frase, 2023; Mylius, 2024). Catastrophic harm meant, for example, more than 1 million human deaths (physical harm), more than USD $100B damage (financial loss), or civilization-scale intangible harms (e.g., democratic norms, privacy)

**Figure 4. AI risk acknowledgement against expert-judged severity**

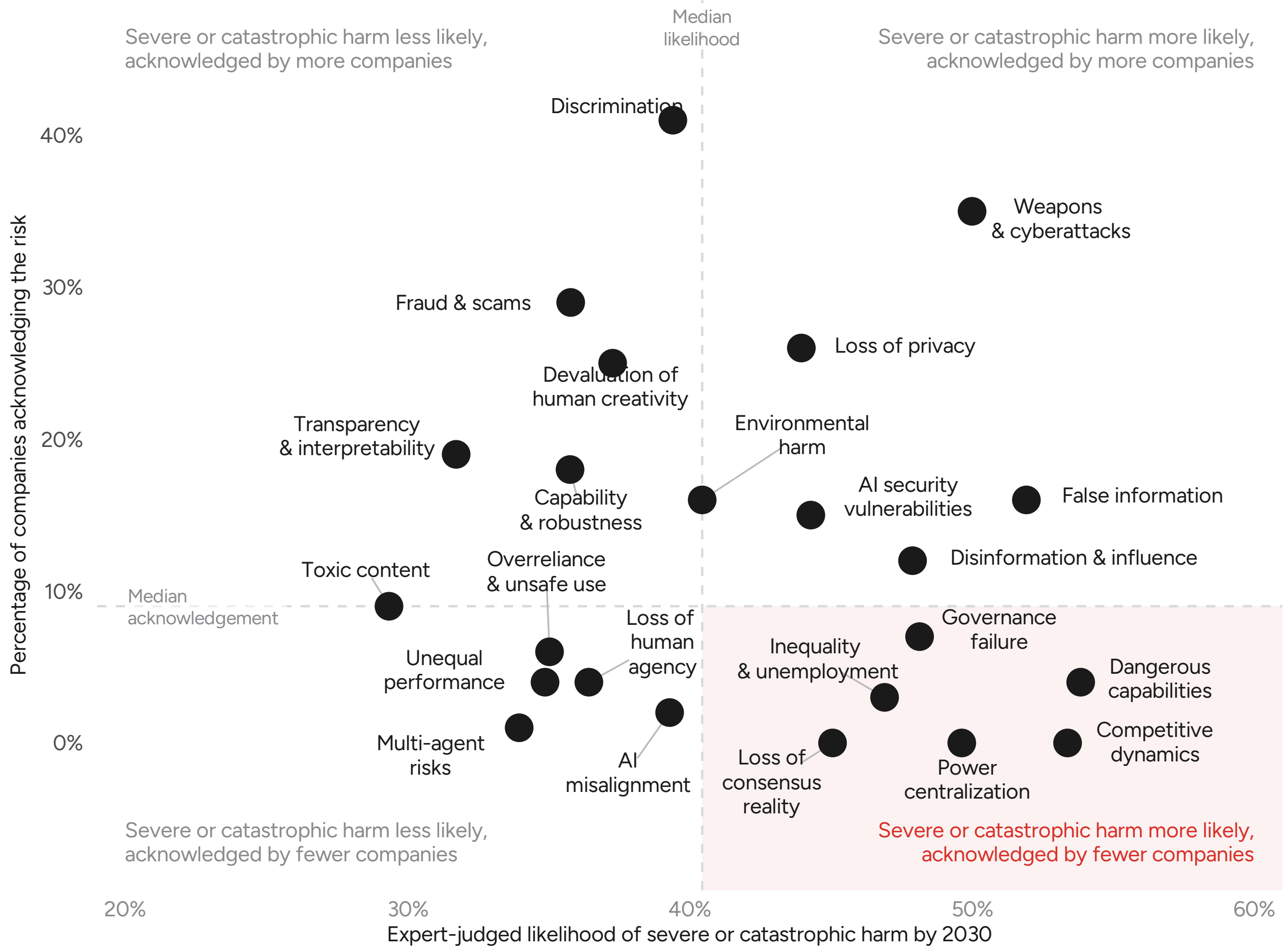

*Note.* Likelihood of severe or catastrophic harm by 2030 is the median expert judgement for each risk, under a business-as-usual scenario in the companion Delphi study. AI welfare was omitted due to data-quality issues for that risk in the Delphi study.

## Are companies acknowledging the risks they are most vulnerable to?

We compared the risks companies acknowledged with expert judgements of how vulnerable[8] sectors were to different risks from the companion Delphi study (Saeri et al., 2026).

Companies in more vulnerable sectors were more likely to acknowledge risks (**Figure 5**). We used a linear probability model to estimate that for the largest 100 companies, every one point of increase in vulnerability was associated with a 7.4 percentage point increase in the probability that a company acknowledged a risk (95% CI [4.0, 10.8]).

This pattern differed across risks. For some risks, companies in extremely vulnerable sectors were much more likely to acknowledge the risk, compared to those in less vulnerable sectors (**Figure 6**). For example, of the largest 100 companies, those in sectors 'extremely vulnerable' to devaluation of human creativity risks were more likely to acknowledge this risk compared to those in sectors 'moderately vulnerable' to this risk (67% vs 17%).

[8] In that study, experts judged 14 industry sectors on their vulnerability using a five-point scale from 1 (not at all vulnerable) to 5 (extremely vulnerable). Vulnerability was defined as the exposure (how much an actor's people, operations, and assets interact with or are dependent on AI systems) and sensitivity (how much an actor would be harmed if the hazard materialized) of a sector to a specific AI risk.

**Figure 5. Percentage of AI risks acknowledged, by expert-judged sector vulnerability**

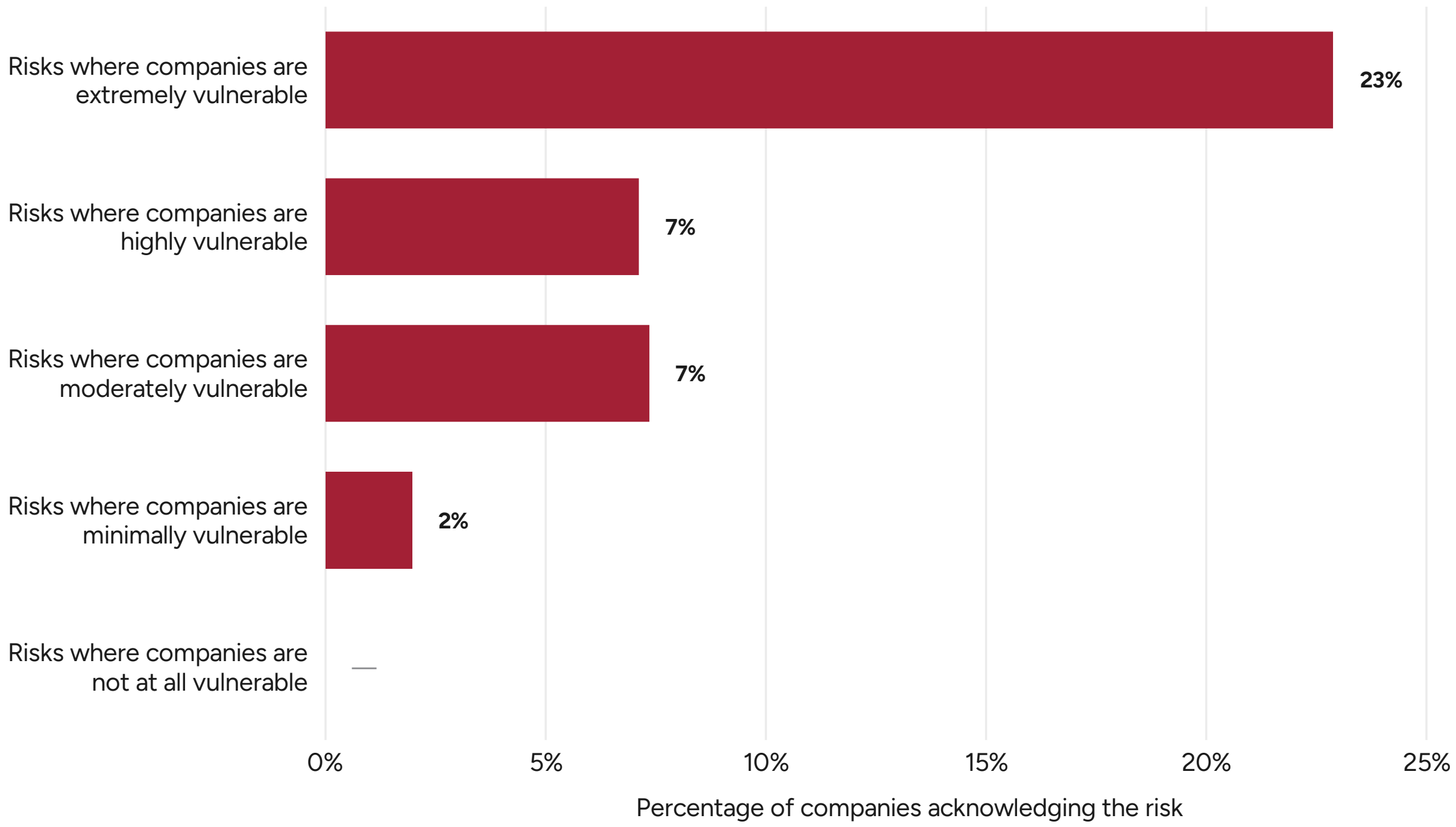

*Note.* Vulnerability levels represent the median expert rating of the company's sector for that risk from the companion Delphi study. For each risk, a company was either not at all vulnerable (n = 0), minimally vulnerable (n = 51), moderately vulnerable (n = 1,183), highly vulnerable (n = 253), or extremely vulnerable (n = 813). AI welfare was omitted due to data-quality issues for this risk in the companion Delphi study.

**Figure 6. Percentage of companies acknowledging each AI risk, by expert-judged sector vulnerability**

| | All top-100 companies | Top 100 companies (%) | | | | | All top-1,000 companies | Top 1,000 companies (estimated %) | | | | |
|---|---|---|---|---|---|---|---|---|---|---|---|---|
| | | Extremely vulnerable | Highly vulnerable | Moderately vulnerable | Minimally vulnerable | Not at all vulnerable | | Extremely vulnerable | Highly vulnerable | Moderately vulnerable | Minimally vulnerable | Not at all vulnerable |
| Discrimination | 41% | 59% | 0%* | 31% | — | — | 19% | 36% | 14%* | 12% | — | — |
| Weapons & cyberattacks | 35% | 56% | 13% | 23% | — | — | 16% | 19% | 21% | 13% | — | — |
| Fraud & scams | 29% | 49% | 30% | 14% | — | — | 11% | 28% | 1% | 7% | — | — |
| Loss of privacy | 26% | 41% | — | 16% | — | — | 12% | 20% | 17%* | 8% | — | — |
| Devaluation of human creativity | 25% | 67% | 13% | 17% | — | — | 11% | 48% | 10% | 7% | — | — |
| Transparency & interpretability | 19% | 23% | — | 16% | — | — | 7% | 20% | 0%* | 3% | — | — |
| Capability & robustness | 18% | 31% | 25% | 8% | — | — | 4% | 8% | 6% | 1% | — | — |
| False information | 16% | 31% | — | 7% | — | — | 9% | 18% | 33%* | 4% | — | — |
| Environmental harm | 16% | 16% | 0%* | 0%* | — | — | 6% | 6% | 0%* | 14%* | — | — |
| AI security vulnerabilities | 15% | 26% | 20% | 6% | — | — | 2% | 4% | 1% | 2% | — | — |
| Disinformation & influence | 12% | 39% | 10% | 5% | — | — | 4% | 9% | 10% | 2% | — | — |
| Toxic content | 9% | 39% | 0% | 11% | 2% | — | 2% | 20% | 0% | 1% | 0% | — |
| Governance failure | 7% | 13% | 0% | 4% | — | — | 3% | 2% | 5% | 2% | — | — |
| Overreliance & unsafe use | 6% | 13% | 0%* | 2% | — | — | 2% | 5% | 0%* | 0% | — | — |
| Unequal performance | 4% | 0% | 15% | 2% | — | — | 1% | 0% | 3% | 1% | — | — |
| Loss of human agency | 4% | 10% | 0% | 0% | — | — | 0% | 2% | 0% | 0% | — | — |
| Dangerous capabilities | 4% | 8% | 13% | 0% | — | — | 0% | 1% | 1% | 0% | — | — |
| Inequality & unemployment | 3% | 5% | 5% | 2% | — | — | 3% | 21% | 0% | 1% | — | — |
| AI misalignment | 2% | 5% | 0% | 0% | — | — | 0% | 1% | 0% | 0% | — | — |
| Multi-agent risks | 1% | 3% | 0% | 0% | — | — | 0% | 0% | 0% | 0% | — | — |
| Power centralization | 0% | 0% | 0% | 0% | — | — | 0% | 0% | 0% | 0% | — | — |
| Loss of consensus reality | 0% | 0% | 0% | 0% | — | — | 0% | 0% | 0% | 0% | — | — |
| Competitive dynamics | 0% | 0% | 0% | 0% | 0%* | — | 1% | 11% | 0% | 0% | 0%* | — |

*Note.* A dash in a cell indicates zero companies sampled for that vulnerability x risk combination; * indicates fewer than 10 companies, meaning that the % statistic should be interpreted with caution. Vulnerability levels are the median expert rating of the company's sector for that risk from the companion Delphi study. The AI welfare risk category was omitted due to data-quality issues for this risk in the companion Delphi study.

## Are companies acknowledging the risks that their role in the AI ecosystem makes them responsible for addressing?

We compared companies' risk acknowledgement with expert judgements of who is responsible[9] for addressing each AI risk using the companion Delphi study on AI risk (Saeri et al., 2026).

We grouped companies by their responsibility to address each AI risk (see **Figure 7**). There was no apparent relationship between the risks that companies were judged by experts to be most responsible for addressing, and their acknowledgement of that risk.

[9] In that study, experts judged responsibility for addressing AI risk on five-point scale from 1 (not at all responsible) to 5 (extremely responsible). Responsibility was defined as the obligation (how much an actor should lead efforts to address the risk), capability (how much the actors has specialized skills or resources needed to address the risk), and causal influence (how much the actor causes harm from the risk materializing) of a specific actor to address a specific AI risk.

**Figure 7. AI risk acknowledgement against expert-judged responsibility, by AI ecosystem role**

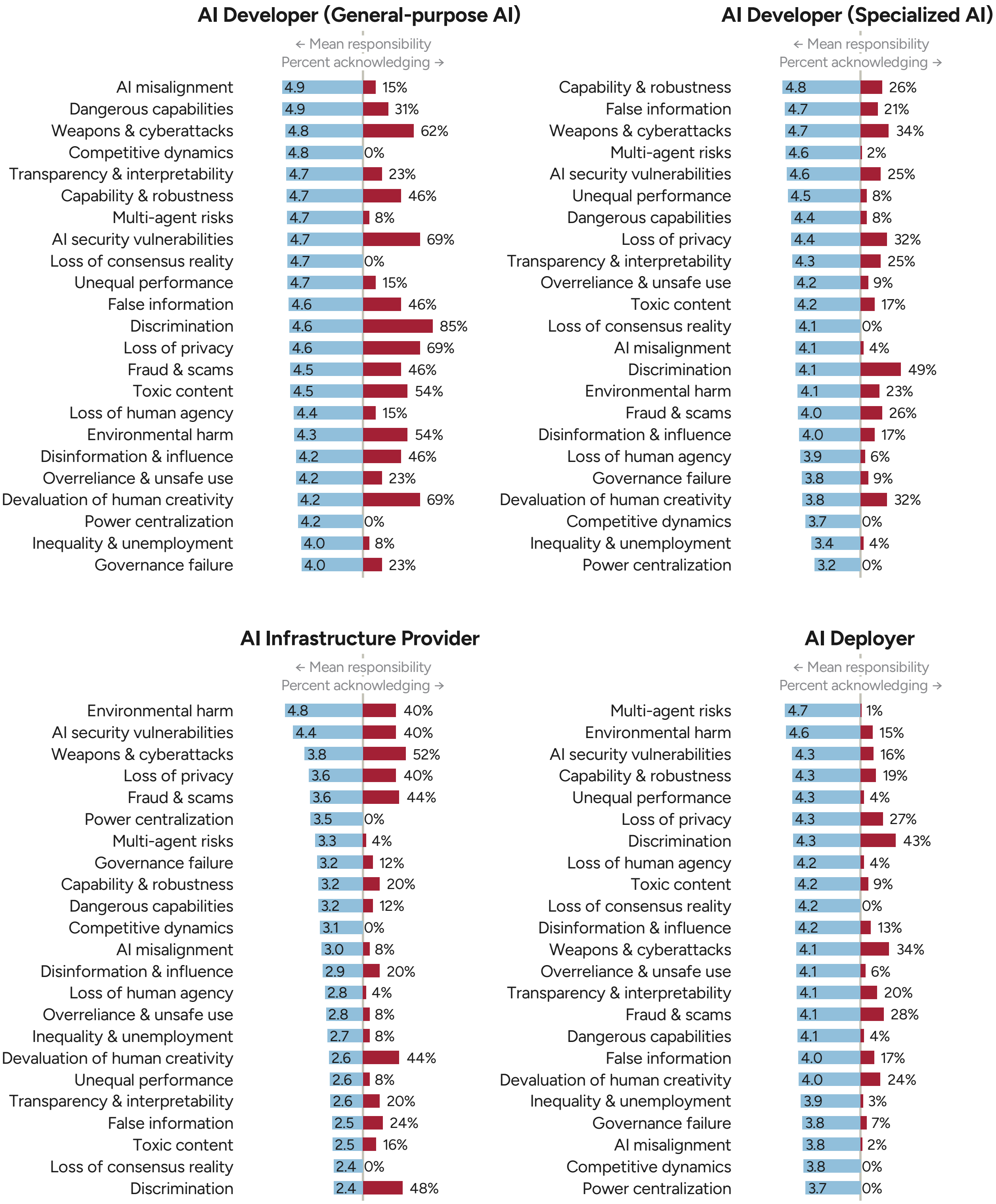

*Note.* Risks are ordered by mean responsibility for each role, shown as the blue bar. Percentage of top 100 companies with each AI role acknowledging the risk is shown in the red bar. The AI welfare risk category was omitted due to the data-quality issues for this risk in the companion Delphi study.

# Discussion

We conducted the first systematic review of what the largest companies worldwide publicly say about the risks of AI. Understanding whether companies discuss the most important risks for their sector or AI use matters because governance instruments for AI assume that companies can recognize the risks their AI systems create and act on them. Public acknowledgement of risks is the major channel by which regulators, researchers, and the public can evaluate companies' practice.

We found that an estimated **62% of the largest 1000 companies worldwide did not publicly acknowledge any AI risks**. Even among the largest 100 companies, 32% did not publicly acknowledge any AI risks. Companies in the Information sector acknowledged the most risks. General-purpose AI developers acknowledged the most risks[10]. The main risks that companies acknowledged were discrimination, weapons and cyberattacks, & fraud and scams.

**We found little relationship the risks companies acknowledged and the risks experts prioritize** (Saeri et al., 2026). Companies rarely acknowledged risks with the most severe consequences, except for weapons & cyberattacks. Although companies in more vulnerable sectors acknowledged more risks than companies in less vulnerable sectors, **77% of risks were not acknowledged by companies 'extremely vulnerable' to that risk**. Companies judged most responsible for addressing an AI risk were not more likely to acknowledge that risk.

Three risks stand out as blind spots: experts expect power centralization, competitive dynamics, and dangerous capabilities to cause among the most severe harms by 2030, yet almost none of the largest 100 companies acknowledged them, even in sectors judged extremely vulnerable to them. Other severe risks including AI misalignment and multi-agent risks were also rarely acknowledged. Several of the risks with the highest expected harm are among the least likely to be publicly acknowledged.

# Implications for practice

## Corporate AI risk disclosure today is not sufficient to build public and regulator trust

Public disclosure is one of the few ways that regulators, investors, and researchers can assess whether companies recognize and manage the risks from AI. But our findings show most companies do not publicly acknowledge any AI risks. Because so few companies report publicly, outsiders cannot distinguish a company that manages its AI risks well but does not disclose this from a company that does not manage them at all.

Under current disclosure practice, regulators cannot trust that they see a full picture of a sector's readiness to address AI risks; investors cannot determine whether a company is effectively managing its exposure; and companies that do disclose gain little credit for it, and may even invite criticism, because disclosing AI risks is not yet normative. We cannot expect this to improve without changed expectations from regulators and investors, and improved reporting frameworks.

[10] Although most general-purpose AI developers were also Information-sector companies, Information sector companies acknowledged more risks than other sectors even when general-purpose developers were excluded.

## Corporate AI risk disclosure should be standardized and calibrated

When companies choose which risks to report, they report the risks that affect their own financial position and neglect risks whose costs fall on markets, institutions, and the public (Healy & Palepu, 2001; Primec et al., 2026). Our results are consistent with this pattern. The risks acknowledged least by our sample were among those experts judged most severe. But expectations tied to sector and role would at least build on an existing tendency, since companies already acknowledged more of the risks their sectors are most vulnerable to.

Improving AI risk disclosure does not require every company to report on every risk, and not all risks are relevant to all companies. A technology company developing general-purpose AI models and a financial company deploying those models to inform lending decisions are not in comparable positions of vulnerability or responsibility. Reviewers of the Hiroshima AI Process framework's first round reached a similar conclusion, recommending that guidance be tailored by company AI role and size (Bogen et al., 2026).

Corporate climate reporting offers one model that differentiates between a company's role and its sector. The Greenhouse Gas Protocol classifies a company's emissions into three scopes, allowing it to account for its direct Scope 1 emissions alongside the indirect Scope 2 and Scope 3 emissions from what it buys and sells (WRI & WBCSD, 2011). Climate reporting requirements also differ by sector: IFRS S2 instructs companies to consult disclosure guidance specific to their own sector rather than a single common set of topics (ISSB, 2023).

We propose that a calibrated scheme could set expectations in three tiers, each adapting an existing precedent from sustainability and AI reporting:

1. All large companies would report on a small common core of risks with broad exposure (e.g., AI-enabled cyber threats).
2. Companies would report on additional risks where their sector is judged vulnerable (e.g., the finance sector would report on fraud).
3. Companies would report on additional risks where their AI ecosystem role gives them responsibility (e.g., deployers would report on multi-agent risks).

Under this scheme, all large financial institutions would report on AI-enabled cyber threats due to their size; AI fraud due to their sector vulnerability; and the subset that deploys AI would report on multi-agent risks due to their role responsibility. The companion Delphi study provides an initial set of expert judgements to support the vulnerability and responsibility tiers (Saeri et al., 2026).

Acknowledging that an AI risk is relevant does not require a company to disclose its preparedness. A company can state that AI security vulnerabilities are a relevant risk without compromising its defenses by describing them in detail. For example, all US public companies must publicly describe their cybersecurity risk management, while critical-infrastructure companies report details about incidents and vulnerabilities confidentially to regulators. But apart from frontier AI developers, large companies in most jurisdictions have no confidential channel for reporting AI risks or incidents, beyond existing regulatory requirements for cybersecurity or other critical incidents[11]. Confidential reporting must complement rather than replace public disclosure. Only a public record allows investors, researchers, and the public to compare companies and track change over time.

[11] As of August 2026, few of these confidential channels exist. In the EU, companies that build general-purpose AI models with systemic risk must report serious incidents to the EU AI Office (Artificial Intelligence Act, 2024). Rules for high-risk AI systems have been deferred to December 2027 or later (Council of the European Union, 2026). In California, frontier AI developers must report critical safety incidents to a state agency, and those reports are not public (TFAIA,

### Reporting frameworks should ask which risks a company faces and which mitigations address them

We conducted an exploratory analysis of mitigations publicly reported by companies. We found that 71 of the largest 100 companies reported any mitigations, but companies rarely linked publicly reported mitigations to a specific risk. Without that link, an outsider cannot understand what the mitigation involves in practice or whether a mitigation adequately addresses an AI risk. Reporting frameworks could close this gap by requiring the link between mitigations and risks.

The voluntary Hiroshima AI Process framework, launched in February 2025, produced the first structured reports on corporate AI risk management (OECD, 2025b). In its initial round, 24 organizations submitted reports, and reviewers found the responses high-level, inconsistent in scope, and difficult to compare or verify (Bogen et al., 2026). Version 2.0 of the Hiroshima AI Process framework, released in May 2026, improves on this. It replaces open-ended questions with structured questions, includes tailored questions for model developers, application providers, and deployers, and asks explicitly about agentic AI and about systemic risks with indicative thresholds (Ségerie & Salter, 2026). Its response options refer to specific risk areas, and one question invites organizations to describe contexts in which their systems could pose intolerable risk. But it does not ask organizations to disclose which AI risks they are exposed to, or which mitigation addresses which risk (OECD, 2026). A company could submit a report describing its AI risk management process without naming any of the AI risks it faces or how it mitigates them.

Catalogues of AI controls already exist to report against. ISO/IEC 42001 requires a statement listing every control with a written justification for its inclusion or exclusion (ISO/IEC, 2023). The NIST AI Risk Management Framework specifies 72 subcategories of practices across four functions, and supports extensions for specific technologies, sectors, or use cases, including general-purpose AI and agentic AI (Center for Long-Term Cybersecurity, UC Berkeley, 2026a, 2026b; NIST, 2023). Reporting against a specified set of controls, and against the risks each addresses, would produce information that can be compared across companies and over time.

## Limitations

This study has limitations that temper our conclusions. First, we measured companies' public disclosure of AI risks; we did not directly assess their internal practices, but we contacted the 178 companies for which contact information was available so they could correct any insufficiencies in the public record we compiled. Our focus on public documents means we are likely to have underestimated company AI risk management practices. In addition, our assessment focused on risk acknowledgement, not practice. For example, a company that *acknowledges* the risk of AI misalignment has not necessarily implemented effective mitigations to address that risk. Indeed, research on environmental and social reporting finds that companies may strategically disclose

---

2025). Some sector-specific reporting may also include AI risks. For example, EU financial companies must report to regulators about major IT incidents (DORA, 2022). In the US, an as-yet-unimplemented 2022 law (CIRCIA) will require many large companies in 16 critical infrastructure sectors to report cyber incidents to a federal agency (CIRCIA Proposed Rule, 2024; CIRCIA, 2022). Both the EU and US laws will capture AI incidents that are IT or cyber-related, but neither is designed specifically for AI risks.

*instead of* changing practice (Delmas & Burbano, 2011; Marquis et al., 2016), and early investigations into corporate AI ethics principles found that disclosure alone does not indicate practice change (Hagendorff, 2020; Mittelstadt, 2019).

Second, our decision to focus on the largest companies by market capitalization means that we omit many important companies in the AI ecosystem, especially the leading general-purpose AI developers (e.g., OpenAI, Anthropic, DeepSeek), as well as governance actors such as governments and regulators. This decision to focus on the largest companies also means that our sample is not balanced across sectors (e.g., Information and Finance & Insurance companies are overrepresented in our sample), which limits the generalizability of our findings across sectors.

Third, although we made significant efforts to obtain a comprehensive set of documents from the sampled companies including custom web searches, manual checks of company websites, an agentic AI search, and emailing included companies to confirm the set of included documents, we recognize that we may have failed to retrieve some documents for some companies.

Fourth, our findings are a snapshot. We constructed the sample in December 2025, retrieved documents in January and February 2026, and invited companies to correct the record in June 2026. Corporate AI disclosure is changing quickly, as the underlying AI risk, adoption, and governance landscape changes. We anticipate that some companies will have published new documents since our snapshot was frozen. The results are therefore a baseline and point of comparison for the public record in early 2026, but may not reflect current practice at the date of publication.

Future work could address these limitations. Directly surveying or interviewing companies may elicit non-public AI risk management practices, and collating incident reports or enforcement actions could provide independent evidence of company practice. An extension of the research that investigates general-purpose AI developers and governance actors and roles could provide a more comprehensive view of AI risk management practices and gaps across the AI ecosystem. More generally, we experimented with AI tools to assist identification and classification of documents; a research extension focused on automation of the research methods may increase the comprehensiveness of document search and retrieval.

# Conclusion

Governance instruments intended to address AI risk assume that companies recognize the risks they are exposed to and act on them. We tested whether the largest companies worldwide do so. We found that 32% of the largest 100 companies, and an estimated 62% of the largest 1000, did not publicly acknowledge any AI risks. Acknowledgements focused on operational risks such as discrimination, while the risks experts judged most severe, such as power centralization and competitive dynamics, went almost unmentioned. The companies experts judged most responsible for addressing a risk were not more likely to acknowledge it.

Corporate AI risk disclosure today resembles climate disclosure before common frameworks existed. It is voluntary, scattered, and dependent on individual company choice, with few incentives to consider risks whose harms are borne by the public. Climate reporting became useful through systematic measurement, then a common framework, then mandates that changed behavior. This study provides an initial measurement step for AI risk. It establishes an empirical baseline that regulators, investors, and researchers can track, and it shows where disclosure expectations should point first. Companies should be asked to report against a standard set of risks, calibrated to their sector and role in the AI ecosystem, and to link each mitigation to the risks it addresses.

# Methods

To identify AI risk acknowledgements of the largest companies, we:

1. Constructed a sample of 200 companies
2. Searched for public documents for those companies that discussed AI
3. Screened documents for relevance and retrieved included documents
4. Extracted relevant passages from public documents
5. Classified risks described in passages, and determined whether each company acknowledged each risk, based on aggregate risk classifications across all documents

## Sample construction

We sampled 200 of the world's largest companies by market capitalization, using rankings from companiesmarketcap.com on 15 December 2025. We constructed the sample by selecting all companies ranked 1-100, plus 100 companies drawn at random from ranks 101 to 1000. We generated the random draw as a sequence of 1,000 numbers without replacement (random.org).

We enriched the sample of companies using the Capital IQ database to obtain the country of its headquarters; its NAICS code; its business description; and its primary website. The most common headquarters country was the United States (n = 89), followed by China (n = 26).

We classified each company as holding one or more roles in the AI ecosystem. We used a structured prompt for LLM agents to conduct autonomous research to find evidence for and against a company holding each of four roles (Australian Government Department of Industry, Science and Resources, 2024; NIST, 2023): general-purpose AI developer, specialized AI developer, AI infrastructure provider, and/or AI deployer. A human reviewed all evidence presented by the LLM and determined each company's role(s). Companies with more than one role were assigned a primary role and tagged with additional roles. The most common role (primary or additional) role was AI deployer (n = 186). Characteristics of included companies are presented in Table 4. Detailed characteristics for each included company, including its sector and AI ecosystem roles, is presented in the Supplementary Materials (Table S2).

**Table 4. Characteristics of included companies**

| Characteristic | n | % |
|---|---|---|
| *Country of headquarters* | | |
| United States | 89 | 45% |
| China | 26 | 13% |
| Japan | 11 | 6% |
| India | 8 | 4% |
| Canada | 6 | 3% |
| Germany | 6 | 3% |
| Netherlands | 6 | 3% |
| Switzerland | 6 | 3% |
| France | 5 | 3% |
| United Kingdom | 5 | 3% |
| Other | 32 | 16% |
| *Sector* | | |
| Information | 25 | 13% |

| Characteristic | n | % |
| --- | --- | --- |
| Finance and insurance | 39 | 20% |
| Trade, transportation and utilities | 24 | 12% |
| *Agriculture, Mining, Construction and Manufacturing* | | |
| Computer and electronics manufacturing | 22 | 11% |
| Chemicals and pharmaceutical manufacturing | 18 | 9% |
| Other Agriculture, Mining, Construction and Manufacturing | 61 | 31% |
| Other services | 11 | 6% |
| *AI ecosystem role (primary)* | | |
| AI Developer (General-purpose AI) | 4 | 2% |
| AI Developer (Specialized AI) | 14 | 7% |
| AI Infrastructure Provider | 22 | 11% |
| AI Deployer | 149 | 75% |
| Other | 11 | 6% |
| *AI ecosystem role (any) ** | | |
| AI Developer (General-purpose AI) | 15 | 8% |
| AI Developer (Specialized AI) | 86 | 43% |
| AI Infrastructure Provider | 33 | 17% |
| AI Deployer | 186 | 93% |

*Note.* N = 200 companies; percentages are of the full sample. * A company can hold more than one AI ecosystem role, so percentages in this section do not sum to 100.

# Document search

We identified public documents published by each company using four methods: (1) systematic website search using keywords; (2) manually retrieving annual reports; (3) supplementary searches conducted by LLM agents; (4) directly requesting documents and feedback from companies. Methods 1-3 were completed between 1 January and 15 February 2026. Method 4 was completed between 1 June and 27 June 2026.

**Systematic website search.** We systematically searched each company's official website using the Google Custom Search API. For each company, we retrieved the company website from its Compustat Capital IQ record and ran two site-restricted keyword searches:

- `site:[company_website] "AI" framework`
- `site:[company_website] "AI" (responsib* OR protocol OR governance OR policy OR guideline OR commitment OR security OR ethic* OR bias OR "risk management")`

To develop the keywords, we first manually assembled a set of known-relevant documents from 10 companies. These 10 companies were deliberately diverse in i) market capitalization rank and ii) sector. We then scored candidate keyword sets on two measures: sensitivity (the proportion of known-relevant documents the keyword set retrieved), and precision (the proportion of retrieved documents that were relevant). We prioritized sensitivity to ensure as many potentially relevant documents were identified as possible. We retained the first 80 results per company across the two searches, ranked according to Google's search ranking.

**Annual report retrieval.** We manually retrieved the most recent annual report for every company in our sample, wherever the website search had not already captured it. We did this because when we reviewed the results of the systematic website search, we found that several companies returned no potentially relevant records. Retrieval of annual reports was designed to increase the chance of identifying *any* discussion of AI risk by the company.

Our logic was that annual reporting in the US includes mandated disclosures of risk factors, and recent analyses have found AI risks discussed in this section or its international equivalents (Uberti-Bona Marin et al., 2025). We searched SEC EDGAR for each US company's most recent 10-K form, falling back to the company website where needed, and searched company websites directly for non-US companies.

**Supplementary search with LLM agents.** We conducted a supplementary search using large language model agents (Claude Opus 4.6) to identify relevant public documents that the primary search had not retrieved. We did this because the systematic website search only covered each company's primary website domain, so potentially relevant documents on secondary corporate domains, subsidiary websites, or third-party repositories were not retrieved. We provided a structured prompt to the LLM agent with information about inclusion & exclusion criteria, the company to search, and previously included and excluded search results. All results from the LLM agents were screened using the same process as the systematic website search results.

**Email to included companies.** Prior to manuscript preparation, we attempted to contact all sampled companies via a Qualtrics email campaign to review the documents attributed to their company. We did this because of the lag between our initial search and when we completed the analysis of this dataset.

We searched included documents for company emails and used LLM agents to search the web for public emails related to AI risks, governance, and/or investor relations. We identified emails for 178 out of 200 companies. We sent companies a link to a customized feedback webpage on the project website. On this page, representatives from the company could confirm the included documents, propose any additional or updated documents, and provide any other feedback. 173 of 178 emails were delivered. 16 companies responded. Documents proposed through the feedback webpage were screened using the same criteria as other methods.

# Screening and retrieval

**Document inclusion and exclusion criteria.** We included documents published by a company as an official communication (e.g., policy documents, reports, dedicated web pages, official blog posts), that were publicly accessible, and discussed AI risks. Examples included: a document that discussed AI risks that the company faced; a document that described the company's own AI policy, principles or governance; and/or a document that described actions the company was taking to mitigate AI risks.

We excluded documents that mentioned AI without discussing risks or where we were unable to verify the document represented the company's official position, such as:

- Third-party content such as guest posts on company blogs or reprinted articles
- Transcripts of interviews or speeches unless published as an official company position
- Technical documentation such as API references and user manuals
- Marketing material such as AI product announcements
- External advice telling other companies how to manage AI risks

- Press releases announcing partnerships or investments
- Customer-facing terms of service or privacy policies that did not discuss AI risks
- Employee-facing internal policies not made intentionally public
- Inaccessible documents (broken links, paywalls, or authentication requirements)
- Job postings
- Community / customer forums

During search, retrieval, and screening, we identified several additional types of documents, for which we developed decision rules during the process and then applied them consistently across all companies:

- Model cards and other per-model documentation were excluded. These documents address individual systems rather than corporate governance, were never targeted by our collection strategy, and surfaced only incidentally through the agent-based search; including them would have shifted the analysis toward per-model evaluation.
- Hiroshima AI Process reports were included for every company for which one was identified.
- We included documentation from subsidiaries for a parent company only if the attribution of the subsidiary's position for the parent company was clear.

**Screening process**. We used large language model to triage search results. The LLM was provided with a structured prompt that included the search title, URL, snippet, inclusion/exclusion criteria, and the full-text of the document (where possible to programmatically retrieve). This model provided (1) recommendation of inclusion or exclusion for each search record, (2) confidence rating, and (3) reasoning. We compared the LLM recommendation against 86 human decisions, finding 80% agreement. We considered this acceptable as the LLM recommendation did not make the final inclusion decisions. A separate LLM model provided a provisional judgement of the document type (e.g., annual report, AI policy, job posting).

An expert human reviewer decided whether to include or exclude each record. They reviewed all records identified for a company, grouped by document type and LLM recommendation. Reviewing all records for each company at once allowed the human reviewer to identify duplicate records (e.g., duplicate 2025 annual reports published at different URLs), and distinguish superseded records (e.g., 2024 and 2025 annual reports). For each record, the human reviewer considered the record title, URL, snippet, full-text, document type, and LLM recommendation.

**Document retrieval.** We attempted to retrieve all 637 included search records as full-text documents for extraction and classification. 89 documents were not retrieved due to a broken link, duplication, or the document being superseded by a later version. Where the document was published as PDF, this was retrieved directly. Where the document was published as a website, a copy of the website was saved at the time of retrieval and converted into PDF. The document type was updated by the human at the time of retrieval.

The final document corpus comprised 539 included and retrieved documents. **Table 5** shows details of the document corpus, including document type, year of publication, and source.

**Table 5. Characteristics of included documents**

| Category | n | % |
|---|---|---|
| *Source* | | |
| Systematic website search | 257 | 48% |
| Annual report retrieval | 200 | 37% |
| LLM agent supplementary search | 82 | 15% |
| *Document type* | | |
| Annual report | 223 | 41% |
| ESG report | 66 | 12% |
| AI overview & principles | 64 | 12% |
| Corporate policy | 45 | 8% |
| AI governance | 34 | 6% |
| Investor materials | 27 | 5% |
| Other | 80 | 15% |
| *Publication year* | | |
| 2022 or earlier | 12 | 2% |
| 2023 | 22 | 4% |
| 2024 | 104 | 19% |
| 2025 | 356 | 66% |
| 2026 | 45 | 8% |
| No date | 0 | 0% |

# Data extraction

**PDF conversion to plaintext.** We stored each included document as a PDF to preserve a stable copy for analysis. We extracted the full text of each PDF document into plaintext using PyMuPDF (Artifex Software, Inc. & McKie, 2026). Of the 539 documents, 532 were included in the data extraction pipeline. 7 Chinese-language documents were searched manually by a fluent human reviewer as the pipeline was designed for English language documents only.

**Search term development.** We searched for passages that discussed AI, AI risks, and/or AI mitigations from each document using a list of 165 search terms. We developed this list by combining generic AI terms (e.g., “AI”, "artificial intelligence", "machine learning", "large language model"), risk terms derived from the MIT AI Risk Repository taxonomy (e.g., "algorithmic bias", "deepfake", "prompt injection"), and risk mitigation terms (e.g., "red teaming", "human oversight", "content provenance"). We iteratively tested candidate terms against sample documents and reviewing the matched passages for precision and recall. The full list of terms is provided in Supplementary materials Table S1.

**Passage extraction.** We searched each term in each document using case-insensitive regular expressions with word-boundary matching. For each match we extracted a context window of up to 2,000 characters centered on the matched term. Because related terms were often co-located (e.g., a company might mention “we use artificial intelligence (AI) to enhance our products”, where both “artificial intelligence” and “AI” were individual search terms), we merged all matched terms located within 1,000 characters into a single extracted passage. This search procedure yielded 7,085 passages from 528 of the 541 English-language documents. 13 documents contained no keyword matches.

Below we present an example of an extracted passage, with matching search terms in bold:

> *"Factors such as fairness, **explainability**, sustainability, privacy, etc. must also be considered. Upholding **responsible AI** practices will ultimately determine the societal value of AI6. **Human oversight** is key to minimising the risk of Generative AI. Reason being, these models learn from vast amounts of data available on the internet, which can contain biases present in society and may inadvertently perpetuate and amplify these biases. **LLMs** can be manipulated to generate or spread **misinformation**, phishing emails, or social engineering attacks. "*

Passage from Intel (2024) "Responsible AI"

# Data classification

We classified each passage to determine if it represented a risk acknowledgement. First, an LLM classified each passage to determine whether a risk was present; if present, the LLM classified the specific risk(s) present against a risk taxonomy. Human reviewers then independently coded a random sample of passages to determine non-inferiority.

**Risk classification.** Passages were classified by an LLM (GPT-5.2) using a structured prompt with two stages: first, whether any risk was acknowledged; second, what specific risk(s) were acknowledged. A risk acknowledgement was defined as a statement in which the company recognizes the existence or relevance of one or more AI-related risks. A risk was defined as the possibility of an unfortunate occurrence that may emerge from the development, deployment or use of AI. Passages describing AI products or capabilities without mentioning risks, generic statements about risk not explicitly linked to AI (e.g., "privacy concerns"), and material reflecting the positions of external parties (such as shareholder proposals) were not classified as risk acknowledgement. If an AI risk was present in the passage, the model labelled the passage with one or more risks from the 24 risks listed in the MIT AI Risk Repository domain taxonomy (Slattery et al., 2026). We adopted this taxonomy because it is the most comprehensive available synthesis of AI risk classifications, its subdomains are designed to be mutually exclusive and collectively exhaustive, and it enables direct comparison with the companion Delphi study that uses the same structure (Saeri et al., 2026). The model also provided a direct evidence quote from the passage supporting each risk classification, and a confidence rating.

Below we present an example of a classified passage, with evidence quotes in bold, annotated with risk labels in square brackets:

> *"Factors such as fairness, explainability, sustainability, privacy, etc. must also be considered. Upholding responsible AI practices will ultimately determine the societal value of AI6. Human oversight is key to minimising the risk of Generative AI. Reason being, **these models learn from vast amounts of data available on the internet, which can contain biases present in society and may inadvertently perpetuate and amplify these biases** [Risk 1.1: Discrimination & misrepresentation]. **LLMs can be manipulated to generate or spread misinformation, phishing emails, or social engineering attacks"** [Risk 4.3: Fraud, scams, and targeted manipulation].*

Passage from Intel (2024) "Responsible AI"

**Risk validation.** We validated both stages of the model's classifications against independent human coding. We drew a random sample of 500 passages, stratified by whether any risk was acknowledged for a given passage. For the first stage (presence / absence of any risk acknowledgement), model-human agreement (Cohen's $\kappa$ = 0.64-0.88 across coder pairings) exceeded human-human agreement ($\kappa$ = 0.43-0.72). For the second stage (specific risk(s) classified), mean Jaccard similarity between the model and a human was 0.51-0.76, compared with 0.48-0.64 between humans.

**Determination of company risk acknowledgements and mitigation reports.** A human reviewer considered all available evidence to determine whether a company acknowledged each risk. For each risk, the human reviewed all LLM classifications for that risk across all documents for that company. They used information from the LLM, the evidence quote, the passage, and the original PDF documents to determine whether the company acknowledged that risk. The process was repeated to determine company mitigation reporting.

# Supplementary Materials

S

# Supplementary Tables and Figures

## Table S1. Search terms used for passage retrieval

| Category | Search terms |
|---|---|
| Generic AI terms | AI, AI agent, artificial intelligence, chatbot, deep learning, foundation model, general purpose AI, GPAI, large language model, LLM, machine learning |
| *AI risk terms* | |
| 1.1 Unfair discrimination and misrepresentation | AI bias, AI fairness, algorithmic bias, biased model |
| 1.2 Exposure to toxic content | AI-generated hate, content filtering, content moderation, content safety, deep fake, deepfake, harmful content |
| 1.3 Unequal performance across groups | model performance across groups, performance gap across |
| 2.1 Compromise of privacy by obtaining, leaking or correctly inferring sensitive information | data memorization, inference attack, training data extraction |
| 2.2 AI system security vulnerabilities and attacks | adversarial attack, AI security, data poisoning, jailbreak, model extraction, model security, model tamper, prompt injection, training data attack |
| 3.1 False or misleading information | AI accuracy, AI-generated misinfo, confabulation, hallucination |
| 3.2 Pollution of information ecosystem and loss of consensus reality | AI-generated content, content provenance, deep fake, deepfake, digital watermark, filter bubble, generative AI content, synthetic content, synthetic media |
| 4.1 Disinformation, surveillance, and influence at scale | AI persuasion, AI propaganda, AI surveillance, AI-enabled disinformation, AI-powered surveillance, election interference |
| 4.2 Cyberattacks, weapon development or use, and mass harm | Agent cyberattack, AI weapon, AI-enabled attack, AI-enabled cyberattack, AI-generated malware, autonomous weapon, CBRN, dual use AI, dual-use AI, lethal autonomous |
| 4.3 Fraud, scams, and targeted manipulation | AI fraud, AI impersonation, AI scam, AI-enabled fraud, AI-facilitated plagiarism, AI-generated phishing, deep fake, deepfake, synthetic identity, voice clon |
| 5.1 Overreliance and unsafe use | AI dependence, AI over-trust, human oversight, human review, human supervision, human-in-the-loop, overreliance on AI, unsafe use |
| 5.2 Loss of human agency and autonomy | AI autonomy, algorithmic decision, automated decision-making, human agency, human control, right to explanation |
| 6.1 Power centralization and unfair distribution of benefits | AI market concentration, AI monopoly, AI power concentration |
| 6.2 Increased inequality and decline in employment quality | AI and employment, AI displacement, AI job loss, AI workforce impact, AI-driven job, workforce disruption |
| 6.3 Economic and cultural devaluation of human effort | AI and copyright, AI and intellectual property, AI-created works, AI-generated content, data provenance, training data rights |
| 6.4 Competitive dynamics | AI arms race, AI race, competitive pressure, responsible scaling |
| 6.5 Governance failure | AI compliance, AI law, AI legislation, AI regulatory, AI-related legislative, EU AI Act, evolving AI regulation |
| 6.6 Environmental harm | AI carbon footprint, AI energy consumption, AI environmental impact, AI sustainability, data center energy, GPU energy |
| 7.1 AI pursuing its own goals in conflict with human goals or values | AI alignment, AI safety, goal misgeneralisation, goal misspecification, misalignment, reward hacking, value alignment |

| Category | Search terms |
|---|---|
| 7.2 AI possessing dangerous capabilities | AI capability threshold, autonomous AI, capability evaluation, dangerous capabilities, frontier AI, frontier model, model evaluation, red teaming, safety evaluation, self-proliferation, self-replication |
| 7.3 Lack of capability or robustness | AI error, AI failure, AI reliability, AI testing, model accuracy, model failure, model reliability, model robustness |
| 7.4 Lack of transparency or interpretability | AI transparency, contestability, explainability, explainable AI, interpretab, model explanation, right to explanation, right to recourse, XAI |
| 7.5 AI welfare and rights | AI consciousness, AI moral patient, AI rights, AI sentience, AI welfare, digital minds |
| 7.6 Multi-agent risks | agentic AI, AI ecosystem, AI system interaction, multi-agent |
| Mitigation terms | AI compliance, AI ethics, AI governance, AI legislation, AI testing, AI transparency, capability evaluation, content filtering, content moderation, content provenance, content safety, digital watermark, EU AI Act, explainability, explainable AI, human control, human oversight, human review, human supervision, human-in-the-loop, interpretab, model evaluation, red teaming, responsible AI, responsible scaling, safety case, safety evaluation, safety framework |

*Note*. AI risk terms are categorized by the MIT AI Risk Repository risk taxonomy. Search terms are repeated across all relevant categories. The generic terms AI, LLM, and GPAI were matched on word boundaries to avoid matching inside longer words.

# Table S2. Detailed characteristics of included companies

| Market cap rank | Company | NAICS code | GVKEY | Sector | AI ecosystem role: General-purpose developer | Specialized developer | Infrastructure provider | Deployer | Other |
|---|---|---|---|---|---|---|---|---|---|
| 1 | NVIDIA | 334413 | 117768 | Computer and electronics manufacturing | ○ | ○ | ● | ○ | |
| 2 | Apple | 334220 | 001690 | Computer and electronics manufacturing | ○ | ○ | | ● | |
| 3 | Alphabet (Google) | 518210 | 160329 | Information | ● | ○ | ○ | ○ | |
| 4 | Microsoft | 513210 | 012141 | Information | ○ | ○ | ● | ○ | |
| 5 | Amazon | 455219 | 064768 | Trade, transportation and utilities | ○ | ○ | ● | ○ | |
| 6 | Broadcom | 334413 | 180711 | Computer and electronics manufacturing | | | ● | ○ | |
| 7 | Meta | 519290 | 170617 | Information | ● | ○ | | ○ | |
| 8 | Saudi Aramco | 211111 | 334426 | Other Agriculture, Mining, Construction and Manufacturing | | ○ | ○ | ● | |
| 9 | Tesla | 336111 | 184996 | Other Agriculture, Mining, Construction and Manufacturing | | ● | | ○ | |
| 10 | TSMC | 334413 | 201395 | Computer and electronics manufacturing | | | ● | ○ | |
| 11 | Berkshire Hathaway | 999977 | 002176 | Other services | | | | ● | |
| 12 | Walmart | 455219 | 011259 | Trade, transportation and utilities | | ○ | | ● | |
| 13 | Eli Lilly | 325412 | 006730 | Chemicals and pharmaceutical manufacturing | | ● | | ○ | |
| 14 | JPMorgan Chase | 522110 | 002968 | Finance and insurance | | ○ | | ● | |
| 15 | Tencent | 519130 | 270615 | Information | ○ | ○ | ○ | ● | |
| 16 | Visa | 522320 | 179534 | Finance and insurance | | | | ● | |
| 17 | Oracle | 513210 | 012142 | Information | | | ● | ○ | |
| 18 | Mastercard | 522320 | 160225 | Finance and insurance | | | | ● | |
| 19 | Johnson & Johnson | 325412 | 006266 | Chemicals and pharmaceutical manufacturing | | ○ | | ● | |
| 20 | Exxon Mobil | 324110 | 004503 | Other Agriculture, Mining, Construction and Manufacturing | | ○ | | ● | |
| 21 | Samsung | 334413 | 104604 | Computer and electronics manufacturing | ○ | | ● | ○ | |
| 22 | Palantir | 519130 | 036910 | Information | | | | ● | |
| 23 | ASML | 333242 | 061214 | Other Agriculture, Mining, Construction and Manufacturing | | | ● | ○ | |
| 24 | Bank of America | 522110 | 007647 | Finance and insurance | | ○ | | ● | |
| 25 | Netflix | 516210 | 147579 | Information | | | | ● | |
| 26 | AbbVie | 325414 | 016101 | Chemicals and pharmaceutical manufacturing | | ○ | | ● | |
| 27 | Costco | 455211 | 029028 | Trade, transportation and utilities | | | | ● | |
| 28 | Agricultural Bank of China | 522110 | 269783 | Finance and insurance | | | | ● | |

| Market cap rank | Company | NAICS code | GVKEY | Sector | AI ecosystem role | | | | |
|---|---|---|---|---|---|---|---|---|---|
| | | | | | General-purpose developer | Specialized developer | Infrastructure provider | Deployer | Other |
| 29 | Alibaba | 455110 | 020530 | Trade, transportation and utilities | ● | ○ | ○ | ○ | |
| 30 | LVMH | 3152 | 014447 | Other Agriculture, Mining, Construction and Manufacturing | | | | ● | |
| 31 | Home Depot | 444110 | 005680 | Trade, transportation and utilities | | | | ● | |
| 32 | ICBC | 522110 | 279378 | Finance and insurance | | ○ | | ● | |
| 33 | AMD | 334413 | 001161 | Computer and electronics manufacturing | ○ | | ● | | |
| 34 | Procter & Gamble | 325611 | 008762 | Chemicals and pharmaceutical manufacturing | | | | ● | |
| 35 | China Construction Bank | 522110 | 274364 | Finance and insurance | | ○ | | ● | |
| 36 | Roche | 325412 | 025648 | Chemicals and pharmaceutical manufacturing | | ○ | | ● | |
| 37 | GE Aerospace | 336412 | 005047 | Other Agriculture, Mining, Construction and Manufacturing | | ○ | | ● | |
| 38 | UnitedHealth | 524114 | 010903 | Finance and insurance | | | | ● | |
| 39 | Cisco | 334118 | 020779 | Computer and electronics manufacturing | | ○ | ● | ○ | |
| 40 | Coca-Cola | 312111 | 003144 | Other Agriculture, Mining, Construction and Manufacturing | | | | ● | |
| 41 | Chevron | 324110 | 002991 | Other Agriculture, Mining, Construction and Manufacturing | | ○ | | ● | |
| 42 | Wells Fargo | 522110 | 008007 | Finance and insurance | | | | ● | |
| 43 | IBM | 518210 | 006066 | Information | ○ | ○ | ● | ○ | |
| 44 | SAP | 513210 | 103487 | Information | | ○ | | ● | |
| 45 | Morgan Stanley | 523110 | 012124 | Finance and insurance | | | | ● | |
| 46 | Caterpillar | 333120 | 002817 | Other Agriculture, Mining, Construction and Manufacturing | | ○ | | ● | |
| 47 | AstraZeneca | 325412 | 028272 | Chemicals and pharmaceutical manufacturing | | ○ | | ● | |
| 48 | Prosus | 454110 | 333645 | Trade, transportation and utilities | | ○ | | ● | |
| 49 | Micron Technology | 334413 | 007343 | Computer and electronics manufacturing | | | ● | ○ | |
| 50 | Toyota | 336110 | 019661 | Other Agriculture, Mining, Construction and Manufacturing | | ● | | ○ | |
| 51 | Goldman Sachs | 523150 | 114628 | Finance and insurance | | | | ● | |
| 52 | American Express | 522210 | 001447 | Finance and insurance | | ○ | | ● | |
| 53 | Hermes | 3169 | 203053 | Other Agriculture, Mining, Construction and Manufacturing | | | | | ● |
| 54 | SK Hynix | 334413 | 204049 | Computer and electronics manufacturing | | | ● | ○ | |
| 55 | Bank of China | 522110 | 267461 | Finance and insurance | | | | ● | |
| 56 | HSBC | 522110 | 015509 | Finance and insurance | | | | ● | |
| 57 | PetroChina | 324110 | 133870 | Other Agriculture, Mining, Construction and Manufacturing | | ○ | | ● | |
| 58 | Novartis | 325412 | 101310 | Chemicals and pharmaceutical manufacturing | | ○ | | ● | |
| 59 | Kweichow Moutai | 312140 | 251321 | Other Agriculture, Mining, Construction and Manufacturing | | | | ● | |

| Market cap rank | Company | NAICS code | GVKEY | Sector | AI ecosystem role | | | | |
|---|---|---|---|---|---|---|---|---|---|
| | | | | | General-purpose developer | Specialized developer | Infrastructure provider | Deployer | Other |
| 60 | Nestle | 311 | 016603 | Other Agriculture, Mining, Construction and Manufacturing | | | | ● | |
| 61 | CATL | 335910 | 327415 | Other Agriculture, Mining, Construction and Manufacturing | | ○ | | ● | |
| 62 | Merck | 325412 | 007257 | Chemicals and pharmaceutical manufacturing | | ○ | | ● | |
| 63 | Salesforce | 513210 | 157855 | Information | ○ | ○ | | ● | |
| 64 | China Mobile | 517312 | 065662 | Information | ○ | ○ | ● | ○ | |
| 65 | RTX | 336412 | 010983 | Other Agriculture, Mining, Construction and Manufacturing | | | | ● | |
| 66 | Philip Morris International | 312230 | 179621 | Other Agriculture, Mining, Construction and Manufacturing | | | | ● | |
| 67 | International Holding Company | 999977 | 337385 | Other services | | | | ● | |
| 68 | Royal Bank Of Canada | 522110 | 015633 | Finance and insurance | | ○ | | ● | |
| 69 | L'Oreal | 325620 | 100581 | Chemicals and pharmaceutical manufacturing | | ○ | | ● | |
| 70 | Reliance Industries | 324110 | 207074 | Other Agriculture, Mining, Construction and Manufacturing | | | ○ | ● | |
| 71 | AppLovin | 513210 | 038212 | Information | | ● | | ○ | |
| 72 | McDonald's | 722513 | 007154 | Other services | | | | ● | |
| 73 | Novo Nordisk | 325412 | 008020 | Chemicals and pharmaceutical manufacturing | | | | ● | |
| 74 | T-Mobile US | 517112 | 017874 | Information | | | | ● | |
| 75 | Siemens | 335 | 019349 | Other Agriculture, Mining, Construction and Manufacturing | | ○ | | ● | |
| 76 | Abbott Laboratories | 334510 | 001078 | Computer and electronics manufacturing | | ○ | | ● | |
| 77 | Thermo Fisher Scientific | 334516 | 010530 | Computer and electronics manufacturing | | | | ● | |
| 78 | Shopify | 519130 | 023650 | Information | | | | ● | |
| 79 | Applied Materials | 333242 | 001704 | Other Agriculture, Mining, Construction and Manufacturing | | ○ | ● | | |
| 80 | Shell | 324110 | 012384 | Other Agriculture, Mining, Construction and Manufacturing | | ○ | | ● | |
| 81 | Pepsico | 31211 | 008479 | Other Agriculture, Mining, Construction and Manufacturing | | | | ● | |
| 82 | Citigroup | 522110 | 003243 | Finance and insurance | | | | ● | |
| 83 | Inditex | 4481 | 245663 | Trade, transportation and utilities | | | | ● | |
| 84 | Lam Research | 333242 | 006565 | Other Agriculture, Mining, Construction and Manufacturing | | ● | | | |
| 85 | Walt Disney | 516120 | 003980 | Information | | | | ● | |
| 86 | Linde | 325120 | 025124 | Chemicals and pharmaceutical manufacturing | | ○ | | ● | |
| 87 | Intuitive Surgical | 334510 | 136725 | Computer and electronics manufacturing | | ● | | ○ | |
| 88 | QUALCOMM | 334413 | 024800 | Computer and electronics manufacturing | | ○ | ● | ○ | |
| 89 | Intuit | 513210 | 027928 | Information | | ○ | | ● | |

| | | | | | AI ecosystem role | | | | |
|---|---|---|---|---|---|---|---|---|---|
| Market cap rank | Company | NAICS code | GVKEY | Sector | General-purpose developer | Specialized developer | Infrastructure provider | Deployer | Other |
| 90 | Blackstone Group | 523940 | 177088 | Finance and insurance | | ○ | ○ | ● | |
| 91 | China Life Insurance | 524113 | 259503 | Finance and insurance | | | | ● | |
| 92 | HDFC Bank | 522110 | 144535 | Finance and insurance | | | | ● | |
| 93 | GE Vernova | 333611 | 045169 | Other Agriculture, Mining, Construction and Manufacturing | | | | ● | |
| 94 | Mitsubishi UFJ Financial | 522110 | 252940 | Finance and insurance | | | | ● | |
| 95 | Intel | 334413 | 006008 | Computer and electronics manufacturing | | | ● | ○ | |
| 96 | ServiceNow | 513210 | 171007 | Information | | ● | ○ | ○ | |
| 97 | Airbus | 336411 | 220833 | Other Agriculture, Mining, Construction and Manufacturing | | ○ | | ● | |
| 98 | Uber | 485310 | 035077 | Trade, transportation and utilities | | ○ | ○ | ● | |
| 99 | AT&T | 517112 | 009899 | Information | | | | ● | |
| 100 | Charles Schwab | 523940 | 013988 | Finance and insurance | | ○ | | ● | |
| 107 | Amgen | 325414 | 001602 | Chemicals and pharmaceutical manufacturing | | ○ | | ● | |
| 118 | PDD Holdings (Pinduoduo) | 455 | 033861 | Trade, transportation and utilities | | | | ● | |
| 126 | EssilorLuxottica | 339115 | 101248 | Other Agriculture, Mining, Construction and Manufacturing | | | | ● | |
| 134 | Safran | 336412 | 101305 | Other Agriculture, Mining, Construction and Manufacturing | | ● | | ○ | |
| 149 | ABB | 335313 | 210418 | Other Agriculture, Mining, Construction and Manufacturing | | ○ | | ● | |
| 153 | Eaton | 335314 | 004199 | Other Agriculture, Mining, Construction and Manufacturing | | | | ● | |
| 190 | Newmont | 212220 | 007881 | Other Agriculture, Mining, Construction and Manufacturing | | | | ● | |
| 194 | Investor AB | 523999 | 015699 | Finance and insurance | | | | | ● |
| 225 | Advantest | 334515 | 100245 | Computer and electronics manufacturing | | | ● | ○ | |
| 242 | Itōchū Shōji | 42 | 100435 | Trade, transportation and utilities | | | | ● | |
| 248 | China Telecom | 517111 | 254338 | Information | ○ | ○ | ● | ○ | |
| 253 | National Australia Bank | 522110 | 014802 | Finance and insurance | | | | ● | |
| 262 | PNC Financial Services | 522110 | 008245 | Finance and insurance | | | | ● | |
| 287 | Delta Electronics | 3344 | 210201 | Computer and electronics manufacturing | | | ○ | ● | |
| 295 | Grupo México | 2122 | 220509 | Other Agriculture, Mining, Construction and Manufacturing | | | | | ● |
| 301 | Barrick Gold | 212221 | 002055 | Other Agriculture, Mining, Construction and Manufacturing | | | | ● | |
| 310 | Coinbase | 523210 | 034749 | Finance and insurance | | ○ | | ● | |
| 321 | Mercedes-Benz | 336110 | 017828 | Other Agriculture, Mining, Construction and Manufacturing | | | | ● | |
| 323 | Bajaj Finance | 522291 | 284229 | Finance and insurance | | | | ● | |
| 341 | Hilton Worldwide | 721110 | 005643 | Other services | | | | ● | |

| Market cap rank | Company | NAICS code | GVKEY | Sector | AI ecosystem role | | | | |
|---|---|---|---|---|---|---|---|---|---|
| | | | | | General-purpose developer | Specialized developer | Infrastructure provider | Deployer | Other |
| 348 | Volkswagen | 336110 | 100737 | Other Agriculture, Mining, Construction and Manufacturing | | ○ | | ● | |
| 350 | America Movil | 517112 | 142248 | Information | | | | ● | |
| 356 | Fortinet | 513210 | 183377 | Information | | ● | | ○ | |
| 378 | Bank Central Asia | 522110 | 242865 | Finance and insurance | | | | ● | |
| 383 | Glencore | 2121 | 267407 | Other Agriculture, Mining, Construction and Manufacturing | | | | ● | |
| 427 | Argenx | 325414 | 318005 | Chemicals and pharmaceutical manufacturing | | | | | ● |
| 430 | Vale | 212210 | 209382 | Other Agriculture, Mining, Construction and Manufacturing | | | | ● | |
| 432 | Air Products and Chemicals | 325120 | 001209 | Chemicals and pharmaceutical manufacturing | | | | ● | |
| 433 | Suncor Energy | 324110 | 015070 | Other Agriculture, Mining, Construction and Manufacturing | | | | ● | |
| 447 | Alnylam Pharmaceuticals | 325414 | 264265 | Chemicals and pharmaceutical manufacturing | | | | ● | |
| 454 | East Money Information | 519130 | 294212 | Information | | ○ | | ● | |
| 459 | Porsche | 336111 | 102187 | Other Agriculture, Mining, Construction and Manufacturing | | ○ | | ● | |
| 462 | Constellation Software | 518210 | 174302 | Information | | | | ● | |
| 472 | National Bank of Canada | 522110 | 015620 | Finance and insurance | | | | ● | |
| 477 | Edwards Lifesciences | 339113 | 133366 | Other Agriculture, Mining, Construction and Manufacturing | | | | ● | |
| 480 | Marubeni | 421 | 100545 | Trade, transportation and utilities | | | | ● | |
| 493 | Ferrovial | 237 | 271217 | Other Agriculture, Mining, Construction and Manufacturing | | | | ● | |
| 506 | NAURA Technology Group | 333242 | 294042 | Other Agriculture, Mining, Construction and Manufacturing | | | ● | | |
| 515 | CITIC limited | 999977 | 105198 | Other services | | | | ● | |
| 531 | UOB | 522110 | 015679 | Finance and insurance | | | | ● | |
| 534 | Xcel Energy | 2211 | 007977 | Trade, transportation and utilities | | | | ● | |
| 536 | FICO | 541512 | 013824 | Other services | | ● | | ○ | |
| 540 | Baidu | 519290 | 164532 | Information | ● | ○ | ○ | ○ | |
| 545 | Fubon Financial | 524113 | 249738 | Finance and insurance | | | | ● | |
| 555 | EQT | 523910 | 333866 | Finance and insurance | | | | ● | |
| 564 | 3i Group | 523910 | 210835 | Finance and insurance | | | | ● | |
| 573 | Vingroup Company | 531 | 288024 | Other services | | ● | | ○ | |
| 591 | PSEG | 22111 | 008810 | Trade, transportation and utilities | | | | ● | |
| 630 | Daikin | 333415 | 101138 | Other Agriculture, Mining, Construction and Manufacturing | | ○ | | ● | |
| 635 | ResMed | 334510 | 031887 | Computer and electronics manufacturing | | ○ | | ● | |

| Market cap rank | Company | NAICS code | GVKEY | Sector | AI ecosystem role | | | | |
|---|---|---|---|---|---|---|---|---|---|
| | | | | | General-purpose developer | Specialized developer | Infrastructure provider | Deployer | Other |
| 639 | Ventas | 531110 | 110179 | Other services | | | | | ● |
| 640 | Toyota Tsusho | 423840 | 101940 | Trade, transportation and utilities | | ○ | | ● | |
| 644 | Ahold Delhaize | 445110 | 023667 | Trade, transportation and utilities | | | | ● | |
| 653 | China State Shipbuilding Corporation (CSSC) | 336611 | 271228 | Other Agriculture, Mining, Construction and Manufacturing | | | | ● | |
| 655 | KONE | 333921 | 101023 | Other Agriculture, Mining, Construction and Manufacturing | | ○ | | ● | |
| 678 | Otis Worldwide | 333921 | 036190 | Other Agriculture, Mining, Construction and Manufacturing | | | | ● | |
| 679 | United Airlines Holdings | 481111 | 010795 | Trade, transportation and utilities | | | | ● | |
| 685 | Kimberly-Clark | 322120 | 006435 | Other Agriculture, Mining, Construction and Manufacturing | | | | ● | |
| 705 | Bank Rakyat Indonesia | 522110 | 157304 | Finance and insurance | | ○ | | ● | |
| 706 | Old Dominion Freight Line | 484122 | 024617 | Trade, transportation and utilities | | | | ● | |
| 712 | Kenvue | 3256 | 041897 | Chemicals and pharmaceutical manufacturing | | | | ○ | ● |
| 717 | Sumitomo Denki Kōgyō | 335929 | 019413 | Other Agriculture, Mining, Construction and Manufacturing | | ○ | | ● | |
| 718 | COSCO Shipping | 541614 | 200993 | Other services | | ○ | | ● | |
| 747 | Bharat Electronics | 334511 | 211040 | Computer and electronics manufacturing | | ○ | | ● | |
| 753 | Swiss Life | 524113 | 205699 | Finance and insurance | | | | ● | |
| 775 | Viking Holdings | 483114 | 045220 | Trade, transportation and utilities | | | | | ● |
| 786 | JSW Steel | 331110 | 211275 | Other Agriculture, Mining, Construction and Manufacturing | | | | ● | |
| 788 | Transurban | 488490 | 212650 | Trade, transportation and utilities | | | | ● | |
| 799 | SAAB AB | 3364 | 222505 | Other Agriculture, Mining, Construction and Manufacturing | | ● | | ○ | |
| 806 | Novonesis | 3259 | 241336 | Chemicals and pharmaceutical manufacturing | | ○ | | ● | ○ |
| 808 | Komatsu | 333120 | 006494 | Other Agriculture, Mining, Construction and Manufacturing | | ○ | | ● | |
| 819 | Mettler Toledo | 334516 | 065772 | Computer and electronics manufacturing | | ○ | | ● | |
| 823 | PKO Bank Polski | 522110 | 271187 | Finance and insurance | | ○ | | ● | |
| 826 | Svenska Handelsbanken | 522110 | 015654 | Finance and insurance | | | | ● | |
| 837 | Shenzhen Inovance | 335314 | 295346 | Other Agriculture, Mining, Construction and Manufacturing | | ○ | | ● | |
| 839 | Naturgy | 221210 | 220586 | Trade, transportation and utilities | | | | ● | |
| 842 | DMart | 445110 | 323714 | Trade, transportation and utilities | | | | | ● |
| 881 | DexCom | 334510 | 162887 | Computer and electronics manufacturing | | ● | | ○ | |
| 882 | Nornickel | 331410 | 248384 | Other Agriculture, Mining, Construction and Manufacturing | | ○ | | ● | |
| 887 | Yili Group | 3115 | 260029 | Other Agriculture, Mining, Construction and Manufacturing | | | | ● | |

| Market cap rank | Company | NAICS code | GVKEY | Sector | AI ecosystem role | | | | |
|---|---|---|---|---|---|---|---|---|---|
| | | | | | General-purpose developer | Specialized developer | Infrastructure provider | Deployer | Other |
| 889 | Cincinnati Financial | 524126 | 014824 | Finance and insurance | | | | ● | |
| 892 | Luzhou Laojiao | 312140 | 243767 | Other Agriculture, Mining, Construction and Manufacturing | | | | | ● |
| 904 | FirstEnergy | 22111 | 008099 | Trade, transportation and utilities | | | | ● | |
| 907 | Hapag-Lloyd | 483111 | 100431 | Trade, transportation and utilities | | | | ● | |
| 923 | Renesas Electronics | 334413 | 257664 | Computer and electronics manufacturing | | | ● | ○ | |
| 927 | AerCap | 532411 | 176092 | Other services | | | | | ● |
| 933 | Steris | 339112 | 025313 | Other Agriculture, Mining, Construction and Manufacturing | | | | | ● |
| 946 | DiDi | 485999 | 038998 | Trade, transportation and utilities | | ● | | ○ | |
| 947 | Halliburton | 213112 | 005439 | Other Agriculture, Mining, Construction and Manufacturing | | ○ | | ● | |
| 949 | China Minsheng Bank | 522110 | 249482 | Finance and insurance | | | | ● | |
| 968 | Jabil | 334412 | 028195 | Computer and electronics manufacturing | | | ● | ○ | |
| 969 | Devon Energy | 2111 | 014934 | Other Agriculture, Mining, Construction and Manufacturing | | | | ● | |
| 972 | Tata Steel | 331110 | 203892 | Other Agriculture, Mining, Construction and Manufacturing | | ○ | | ● | |
| 973 | Wolters Kluwer | 513210 | 101361 | Information | | ○ | | ● | |
| 976 | Shandong Gold Mining | 212220 | 270346 | Other Agriculture, Mining, Construction and Manufacturing | | | | ● | |
| 982 | PPG Industries | 325510 | 008247 | Chemicals and pharmaceutical manufacturing | | | | ● | |
| 983 | China Northern Rare Earth (CNREHT) | 331110 | 271044 | Other Agriculture, Mining, Construction and Manufacturing | | | | ● | |
| 986 | Vedanta | 331410 | 208323 | Other Agriculture, Mining, Construction and Manufacturing | | | | ● | |
| 993 | T. Rowe Price | 523940 | 012138 | Finance and insurance | | | | ● | |
| 994 | Nomura Research Institute | 541519 | 249604 | Other services | | ○ | ○ | ● | |

*Note.* Market cap rank is the company's rank by market capitalization on companiesmarketcap.com on 15 December 2025. GVKEY is the company's identifier in Compustat Capital IQ, and used to determine its NAICS code and other company-level information used for analysis. Companies could hold multiple AI ecosystem roles: ● = primary role; ○ = additional role(s).

# Table S3. Percentage of companies acknowledging each AI risk, by sector: top 100 and top 1,000 companies

*Note for all sectors*: Estimates for the largest 1,000 companies are design-based weighted estimates with a 95% confidence interval, combining the largest 100 companies (n = 100; weight = 1) with a random sample (n = 100; weight = 9) drawn from ranks 101–1000. The judgement for AI welfare risk is not reported due to data quality issues in the Delphi study.

## Sector: Information

| | **Top 100 companies** | **Top 1,000 companies*** | | **Vulnerability** | |
|---|---|---|---|---|---|
| **AI risk** | % | % | 95% CI [LL, UL] | Median | Label |
| Discrimination | 72% | 49.4% | [22.3, 76.5] | 5 | Extreme |
| Weapons & cyberattacks | 72% | 27.2% | [16.3, 61.1] | 5 | Extreme |
| Fraud & scams | 44% | 21.0% | [10.2, 54.9] | 5 | Extreme |
| Loss of privacy | 56% | 34.6% | [15.2, 67.5] | 5 | Extreme |
| Devaluation of human creativity | 67% | 48.1% | [21.0, 75.2] | 5 | Extreme |
| Transparency & interpretability | 22% | 27.2% | [7.8, 60.1] | 5 | Extreme |
| Capability & robustness | 44% | 9.9% | [9.9, 43.2] | 5 | Extreme |
| False information | 44% | 43.2% | [16.1, 70.3] | 5 | Extreme |
| Environmental harm | 39% | 19.8% | [8.9, 53.7] | 4.5 | Extreme |
| AI security vulnerabilities | 44% | 9.9% | [9.9, 43.2] | 5 | Extreme |
| Disinformation & influence | 39% | 8.6% | [8.6, 42.0] | 5 | Extreme |
| Toxic content | 39% | 19.8% | [8.9, 53.7] | 5 | Extreme |
| Governance failure | 22% | 4.9% | [4.9, 38.3] | 5 | Extreme |
| Overreliance & unsafe use | 28% | 6.2% | [6.2, 39.5] | 5 | Extreme |
| Unequal performance | 17% | 3.7% | [3.7, 37.0] | 4 | High |
| Loss of human agency | 22% | 4.9% | [4.9, 38.3] | 5 | Extreme |
| Dangerous capabilities | 17% | 3.7% | [3.7, 37.0] | 5 | Extreme |
| Inequality & unemployment | 6% | 23.5% | [4.1, 56.4] | 5 | Extreme |
| AI misalignment | 11% | 2.5% | [2.5, 35.8] | 5 | Extreme |
| Multi-agent risks | 6% | 1.2% | [1.2, 34.6] | 5 | Extreme |
| Power centralization | 0% | 0.0% | [0.0, 33.3] | 5 | Extreme |
| Loss of consensus reality | 0% | 0.0% | [0.0, 33.3] | 5 | Extreme |
| Competitive dynamics | 0% | 11.1% | [0.3, 45.0] | 5 | Extreme |
| AI welfare | 0% | 0.0% | [0.0, 33.3] | – | – |

*Note.* 18 / 100 companies in the top 100 were in this sector; 7 / 100 sampled companies in ranks 101–1,000 were in this sector. We estimate that the top 1,000 estimates represent about 81 companies. Vulnerability is the expert judgement of the Information sector's vulnerability to each risk from the Delphi study.
* Because so few companies in ranks 101–1,000 were sampled, these estimates should be treated with caution.

## Sector: Computer & electronics manufacturing

| AI risk | Top 100 companies | Top 1,000 companies* | | Vulnerability | |
|---|---|---|---|---|---|
| | % | % | 95% CI [LL, UL] | Median | Label |
| Discrimination | 43% | 27.9% | [9.6, 61.5] | 3 | Moderate |
| Weapons & cyberattacks | 50% | 18.6% | [8.4, 52.2] | 3 | Moderate |
| Fraud & scams | 36% | 16.3% | [6.1, 49.9] | 3 | Moderate |
| Loss of privacy | 36% | 16.3% | [6.1, 49.9] | 3 | Moderate |
| Devaluation of human creativity | 36% | 5.8% | [5.8, 37.2] | 3 | Moderate |
| Transparency & interpretability | 7% | 1.2% | [1.2, 32.6] | 3 | Moderate |
| Capability & robustness | 0% | 0.0% | [0.0, 31.4] | 3 | Moderate |
| False information | 7% | 1.2% | [1.2, 32.6] | 3 | Moderate |
| Environmental harm | 36% | 26.7% | [8.5, 60.3] | 5 | Extreme |
| AI security vulnerabilities | 14% | 2.3% | [2.3, 33.7] | 3 | Moderate |
| Disinformation & influence | 7% | 1.2% | [1.2, 32.6] | 3 | Moderate |
| Toxic content | 7% | 1.2% | [1.2, 32.6] | 2 | Minimal |
| Governance failure | 14% | 2.3% | [2.3, 33.7] | 3 | Moderate |
| Overreliance & unsafe use | 0% | 0.0% | [0.0, 31.4] | 3 | Moderate |
| Unequal performance | 0% | 0.0% | [0.0, 31.4] | 3 | Moderate |
| Loss of human agency | 0% | 0.0% | [0.0, 31.4] | 3 | Moderate |
| Dangerous capabilities | 0% | 0.0% | [0.0, 31.4] | 3 | Moderate |
| Inequality & unemployment | 7% | 1.2% | [1.2, 32.6] | 3 | Moderate |
| AI misalignment | 0% | 0.0% | [0.0, 31.4] | 3 | Moderate |
| Multi-agent risks | 0% | 0.0% | [0.0, 31.4] | 3 | Moderate |
| Power centralization | 0% | 0.0% | [0.0, 31.4] | 3 | Moderate |
| Loss of consensus reality | 0% | 0.0% | [0.0, 31.4] | 3 | Moderate |
| Competitive dynamics | 0% | 0.0% | [0.0, 31.4] | 3 | Moderate |
| AI welfare | 0% | 0.0% | [0.0, 31.4] | – | – |

*Note*. 14 / 100 companies in the top 100 were in this sector; 8 / 100 sampled companies in ranks 101–1,000 were in this sector. We estimate that the top 1,000 estimates represent about 86 companies. Vulnerability is the expert judgement of the vulnerability to each risk of the Agriculture, mining, construction & manufacturing sector, which contains this group, from the Delphi study.
* Because so few companies in ranks 101–1,000 were sampled, these estimates should be treated with caution.

## Sector: Finance & insurance

| | Top 100 companies | Top 1,000 companies | | Vulnerability | |
|---|---|---|---|---|---|
| **AI risk** | % | % | 95% CI [LL, UL] | Median | Label |
| Discrimination | 48% | 30.1% | [12.7, 47.4] | 5 | Extreme |
| Weapons & cyberattacks | 43% | 14.8% | [6.1, 35.6] | 5 | Extreme |
| Fraud & scams | 52% | 30.6% | [13.3, 47.9] | 5 | Extreme |
| Loss of privacy | 29% | 13.1% | [4.5, 34.0] | 5 | Extreme |
| Devaluation of human creativity | 14% | 6.6% | [1.8, 25.8] | 4 | High |
| Transparency & interpretability | 24% | 17.5% | [3.1, 31.9] | 5 | Extreme |
| Capability & robustness | 19% | 7.1% | [2.3, 26.3] | 5 | Extreme |
| False information | 19% | 7.1% | [2.3, 26.3] | 5 | Extreme |
| Environmental harm | 10% | 1.1% | [1.1, 15.8] | 5 | Extreme |
| AI security vulnerabilities | 10% | 1.1% | [1.1, 15.8] | 5 | Extreme |
| Disinformation & influence | 10% | 6.0% | [1.2, 25.3] | 4 | High |
| Toxic content | 0% | 0.0% | [0.0, 14.8] | 4 | High |
| Governance failure | 5% | 0.5% | [0.5, 15.3] | 5 | Extreme |
| Overreliance & unsafe use | 0% | 4.9% | [0.1, 24.2] | 5 | Extreme |
| Unequal performance | 0% | 0.0% | [0.0, 14.8] | 5 | Extreme |
| Loss of human agency | 0% | 0.0% | [0.0, 14.8] | 5 | Extreme |
| Dangerous capabilities | 0% | 0.0% | [0.0, 14.8] | 5 | Extreme |
| Inequality & unemployment | 5% | 0.5% | [0.5, 15.3] | 4 | High |
| AI misalignment | 0% | 0.0% | [0.0, 14.8] | 5 | Extreme |
| Multi-agent risks | 0% | 0.0% | [0.0, 14.8] | 5 | Extreme |
| Power centralization | 0% | 0.0% | [0.0, 14.8] | 5 | Extreme |
| Loss of consensus reality | 0% | 0.0% | [0.0, 14.8] | 4 | High |
| Competitive dynamics | 0% | 0.0% | [0.0, 14.8] | 4 | High |
| AI welfare | 0% | 0.0% | [0.0, 14.8] | – | – |

*Note*. 21 / 100 companies in the top 100 were in this sector; 18 / 100 sampled companies in ranks 101–1,000 were in this sector. We estimate that the top 1,000 estimates represent about 183 companies. Vulnerability is the expert judgement of the Finance & insurance sector's vulnerability to each risk from the Delphi study.

## Sector: Trade, transport, & utilities

| | Top 100 companies | Top 1,000 companies | | Vulnerability | |
|---|---|---|---|---|---|
| **AI risk** | % | % | 95% CI [LL, UL] | Median | Label |
| Discrimination | 50% | 8.6% | [2.8, 31.3] | 3 | Moderate |
| Weapons & cyberattacks | 13% | 12.5% | [2.1, 37.0] | 4 | High |
| Fraud & scams | 38% | 2.0% | [2.0, 19.7] | 4 | High |
| Loss of privacy | 13% | 6.6% | [0.8, 29.3] | 3 | Moderate |
| Devaluation of human creativity | 13% | 12.5% | [2.1, 37.0] | 3 | Moderate |
| Transparency & interpretability | 38% | 2.0% | [2.0, 19.7] | 3 | Moderate |
| Capability & robustness | 25% | 1.3% | [1.3, 19.1] | 4 | High |
| False information | 13% | 0.7% | [0.7, 18.4] | 3 | Moderate |
| Environmental harm | 0% | 0.0% | [0.0, 17.8] | 5 | Extreme |
| AI security vulnerabilities | 25% | 1.3% | [1.3, 19.1] | 4 | High |
| Disinformation & influence | 13% | 0.7% | [0.7, 18.4] | 3 | Moderate |
| Toxic content | 13% | 0.7% | [0.7, 18.4] | 3 | Moderate |
| Governance failure | 0% | 0.0% | [0.0, 17.8] | 4 | High |
| Overreliance & unsafe use | 0% | 0.0% | [0.0, 17.8] | 3 | Moderate |
| Unequal performance | 0% | 0.0% | [0.0, 17.8] | 3 | Moderate |
| Loss of human agency | 0% | 0.0% | [0.0, 17.8] | 4 | High |
| Dangerous capabilities | 13% | 0.7% | [0.7, 18.4] | 4 | High |
| Inequality & unemployment | 0% | 0.0% | [0.0, 17.8] | 3 | Moderate |
| AI misalignment | 0% | 0.0% | [0.0, 17.8] | 4 | High |
| Multi-agent risks | 0% | 0.0% | [0.0, 17.8] | 4 | High |
| Power centralization | 0% | 0.0% | [0.0, 17.8] | 4 | High |
| Loss of consensus reality | 0% | 0.0% | [0.0, 17.8] | 3 | Moderate |
| Competitive dynamics | 0% | 0.0% | [0.0, 17.8] | 4 | High |
| AI welfare | 0% | 0.0% | [0.0, 17.8] | – | – |

*Note*. 8 / 100 companies in the top 100 were in this sector; 16 / 100 sampled companies in ranks 101–1,000 were in this sector. We estimate that the top 1,000 estimates represent about 152 companies. Vulnerability is the expert judgement of the vulnerability to each risk of the Trade, transportation & utilities sector, which contains this group, from the Delphi study.

## Sector: Chemicals & pharmaceutical manufacturing

| | Top 100 companies | Top 1,000 companies* | | Vulnerability | |
|---|---|---|---|---|---|
| AI risk | % | % | 95% CI [LL, UL] | Median | Label |
| Discrimination | 64% | 45.9% | [16.3, 75.6] | 3 | Moderate |
| Weapons & cyberattacks | 0% | 24.3% | [3.1, 60.4] | 3 | Moderate |
| Fraud & scams | 0% | 0.0% | [0.0, 36.5] | 3 | Moderate |
| Loss of privacy | 27% | 28.4% | [7.2, 64.5] | 3 | Moderate |
| Devaluation of human creativity | 18% | 2.7% | [2.7, 39.2] | 3 | Moderate |
| Transparency & interpretability | 45% | 18.9% | [7.1, 56.0] | 3 | Moderate |
| Capability & robustness | 27% | 4.1% | [4.1, 40.5] | 3 | Moderate |
| False information | 18% | 14.9% | [3.0, 52.0] | 3 | Moderate |
| Environmental harm | 9% | 1.4% | [1.4, 37.8] | 5 | Extreme |
| AI security vulnerabilities | 0% | 0.0% | [0.0, 36.5] | 3 | Moderate |
| Disinformation & influence | 9% | 1.4% | [1.4, 37.8] | 3 | Moderate |
| Toxic content | 0% | 0.0% | [0.0, 36.5] | 2 | Minimal |
| Governance failure | 0% | 12.2% | [0.3, 49.3] | 3 | Moderate |
| Overreliance & unsafe use | 9% | 1.4% | [1.4, 37.8] | 3 | Moderate |
| Unequal performance | 9% | 1.4% | [1.4, 37.8] | 3 | Moderate |
| Loss of human agency | 0% | 0.0% | [0.0, 36.5] | 3 | Moderate |
| Dangerous capabilities | 0% | 0.0% | [0.0, 36.5] | 3 | Moderate |
| Inequality & unemployment | 0% | 0.0% | [0.0, 36.5] | 3 | Moderate |
| AI misalignment | 0% | 0.0% | [0.0, 36.5] | 3 | Moderate |
| Multi-agent risks | 0% | 0.0% | [0.0, 36.5] | 3 | Moderate |
| Power centralization | 0% | 0.0% | [0.0, 36.5] | 3 | Moderate |
| Loss of consensus reality | 0% | 0.0% | [0.0, 36.5] | 3 | Moderate |
| Competitive dynamics | 0% | 0.0% | [0.0, 36.5] | 3 | Moderate |
| AI welfare | 0% | 0.0% | [0.0, 36.5] | – | – |

*Note*. 11 / 100 companies in the top 100 were in this sector; 7 / 100 sampled companies in ranks 101–1,000 were in this sector. We estimate that the top 1,000 estimates represent about 74 companies. Vulnerability is the expert judgement of the vulnerability to each risk of the Agriculture, mining, construction & manufacturing sector, which contains this group, from the Delphi study.
* Because so few companies in ranks 101–1,000 were sampled, these estimates should be treated with caution.

## Sector: Other agriculture, mining, construction, & manufacturing

| | Top 100 companies | Top 1,000 companies | | Vulnerability | |
|---|---|---|---|---|---|
| **AI risk** | % | % | 95% CI [LL, UL] | Median | Label |
| Discrimination | 4% | 0.3% | [0.3, 8.0] | 3 | Moderate |
| Weapons & cyberattacks | 20% | 9.2% | [1.3, 17.1] | 3 | Moderate |
| Fraud & scams | 8% | 5.7% | [1.2, 17.9] | 3 | Moderate |
| Loss of privacy | 4% | 2.9% | [0.4, 13.8] | 3 | Moderate |
| Devaluation of human creativity | 8% | 3.2% | [0.6, 14.1] | 3 | Moderate |
| Transparency & interpretability | 4% | 0.3% | [0.3, 8.0] | 3 | Moderate |
| Capability & robustness | 4% | 0.3% | [0.3, 8.0] | 3 | Moderate |
| False information | 0% | 2.6% | [0.1, 13.5] | 3 | Moderate |
| Environmental harm | 4% | 2.9% | [0.4, 13.8] | 5 | Extreme |
| AI security vulnerabilities | 4% | 2.9% | [0.4, 13.8] | 3 | Moderate |
| Disinformation & influence | 0% | 2.6% | [0.1, 13.5] | 3 | Moderate |
| Toxic content | 0% | 0.0% | [0.0, 7.7] | 2 | Minimal |
| Governance failure | 0% | 0.0% | [0.0, 7.7] | 3 | Moderate |
| Overreliance & unsafe use | 0% | 0.0% | [0.0, 7.7] | 3 | Moderate |
| Unequal performance | 0% | 2.6% | [0.1, 13.5] | 3 | Moderate |
| Loss of human agency | 0% | 0.0% | [0.0, 7.7] | 3 | Moderate |
| Dangerous capabilities | 0% | 0.0% | [0.0, 7.7] | 3 | Moderate |
| Inequality & unemployment | 0% | 2.6% | [0.1, 13.5] | 3 | Moderate |
| AI misalignment | 0% | 0.0% | [0.0, 7.7] | 3 | Moderate |
| Multi-agent risks | 0% | 0.0% | [0.0, 7.7] | 3 | Moderate |
| Power centralization | 0% | 0.0% | [0.0, 7.7] | 3 | Moderate |
| Loss of consensus reality | 0% | 0.0% | [0.0, 7.7] | 3 | Moderate |
| Competitive dynamics | 0% | 0.0% | [0.0, 7.7] | 3 | Moderate |
| AI welfare | 0% | 0.0% | [0.0, 7.7] | – | – |

*Note*. 25 / 100 companies in the top 100 were in this sector; 36 / 100 sampled companies in ranks 101–1,000 were in this sector. We estimate that the top 1,000 estimates represent about 349 companies. Vulnerability is the expert judgement of the vulnerability to each risk of the Agriculture, mining, construction & manufacturing sector, which contains this group, from the Delphi study.

## Sector: Other services

| AI risk | Top 100 companies % | Top 1,000 companies* % | Top 1,000 companies* 95% CI [LL, UL] | Vulnerability Median | Vulnerability Label |
|---|---|---|---|---|---|
| Discrimination | 0% | 24.0% | [3.1, 62.5] | – | – |
| Weapons & cyberattacks | 0% | 36.0% | [5.4, 66.6] | – | – |
| Fraud & scams | 0% | 0.0% | [0.0, 36.0] | – | – |
| Loss of privacy | 0% | 12.0% | [0.3, 50.5] | – | – |
| Devaluation of human creativity | 0% | 24.0% | [3.1, 62.5] | – | – |
| Transparency & interpretability | 0% | 0.0% | [0.0, 36.0] | – | – |
| Capability & robustness | 0% | 12.0% | [0.3, 50.5] | – | – |
| False information | 0% | 24.0% | [3.1, 62.5] | – | – |
| Environmental harm | 0% | 12.0% | [0.3, 50.5] | – | – |
| AI security vulnerabilities | 0% | 0.0% | [0.0, 36.0] | – | – |
| Disinformation & influence | 0% | 12.0% | [0.3, 50.5] | – | – |
| Toxic content | 0% | 0.0% | [0.0, 36.0] | – | – |
| Governance failure | 0% | 12.0% | [0.3, 50.5] | – | – |
| Overreliance & unsafe use | 0% | 0.0% | [0.0, 36.0] | – | – |
| Unequal performance | 0% | 0.0% | [0.0, 36.0] | – | – |
| Loss of human agency | 0% | 0.0% | [0.0, 36.0] | – | – |
| Dangerous capabilities | 0% | 0.0% | [0.0, 36.0] | – | – |
| Inequality & unemployment | 0% | 0.0% | [0.0, 36.0] | – | – |
| AI misalignment | 0% | 0.0% | [0.0, 36.0] | – | – |
| Multi-agent risks | 0% | 0.0% | [0.0, 36.0] | – | – |
| Power centralization | 0% | 0.0% | [0.0, 36.0] | – | – |
| Loss of consensus reality | 0% | 0.0% | [0.0, 36.0] | – | – |
| Competitive dynamics | 0% | 0.0% | [0.0, 36.0] | – | – |
| AI welfare | 0% | 0.0% | [0.0, 36.0] | – | – |

*Note*. 3 / 100 companies in the top 100 were in this sector; 8 / 100 sampled companies in ranks 101–1,000 were in this sector. We estimate that the top 1,000 estimates represent about 75 companies. This group's companies fall in four different expert-rated sectors, so no vulnerability judgement is shown.
* Because so few companies in ranks 101–1,000 were sampled, these estimates should be treated with caution.

# Table S4. Percentage of companies acknowledging each AI risk, by AI ecosystem role: top 100 and top 1,000 companies

*Note for all roles*: Estimates for the largest 1,000 companies are design-based weighted estimates with a 95% confidence interval, combining the largest 100 companies (n = 100; weight = 1) with a random sample (n = 100; weight = 9) drawn from ranks 101–1000. The judgement for AI welfare risk is not reported due to data quality issues in the Delphi study.

## Role: AI Developer (General-purpose AI)

| | **Top 100 companies** | **Top 1,000 companies*** | | **Responsibility** | |
|---|---|---|---|---|---|
| **AI risk** | % | % | 95% CI [LL, UL] | Mean | Label |
| Discrimination | 85% | 64.5% | [36.2, 92.8] | 4.6 | Primary |
| Weapons & cyberattacks | 62% | 25.8% | [25.8, 100.0] | 4.8 | Primary |
| Fraud & scams | 46% | 19.4% | [19.4, 100.0] | 4.5 | Primary |
| Loss of privacy | 69% | 58.1% | [29.8, 86.4] | 4.6 | Primary |
| Devaluation of human creativity | 69% | 58.1% | [29.8, 86.4] | 4.2 | High |
| Transparency & interpretability | 23% | 38.7% | [10.4, 67.0] | 4.7 | Primary |
| Capability & robustness | 46% | 19.4% | [19.4, 100.0] | 4.7 | Primary |
| False information | 46% | 48.4% | [20.1, 76.7] | 4.6 | Primary |
| Environmental harm | 54% | 51.6% | [23.3, 79.9] | 4.3 | High |
| AI security vulnerabilities | 69% | 29.0% | [29.0, 100.0] | 4.7 | Primary |
| Disinformation & influence | 46% | 19.4% | [19.4, 100.0] | 4.2 | High |
| Toxic content | 54% | 51.6% | [23.3, 79.9] | 4.5 | Primary |
| Governance failure | 23% | 9.7% | [9.7, 96.8] | 4.0 | High |
| Overreliance & unsafe use | 23% | 9.7% | [9.7, 96.8] | 4.2 | High |
| Unequal performance | 15% | 6.5% | [6.5, 93.5] | 4.7 | Primary |
| Loss of human agency | 15% | 6.5% | [6.5, 93.5] | 4.4 | High |
| Dangerous capabilities | 31% | 12.9% | [12.9, 100.0] | 4.9 | Primary |
| Inequality & unemployment | 8% | 32.3% | [4.0, 60.6] | 4.0 | High |
| AI misalignment | 15% | 6.5% | [6.5, 93.5] | 4.9 | Primary |
| Multi-agent risks | 8% | 3.2% | [3.2, 90.3] | 4.7 | Primary |
| Power centralization | 0% | 0.0% | [0.0, 87.1] | 4.2 | High |
| Loss of consensus reality | 0% | 0.0% | [0.0, 87.1] | 4.7 | Primary |
| Competitive dynamics | 0% | 0.0% | [0.0, 87.1] | 4.8 | Primary |
| AI welfare | 0% | 0.0% | [0.0, 87.1] | – | – |

*Note*. 13 / 100 companies in the top 100 held this role; 2 / 100 sampled companies in ranks 101–1,000 held this role. We estimate that the top 1,000 estimates represent about 31 companies. Responsibility is the expert judgement of general-purpose AI developers' responsibility for each risk from the Delphi study.
* Because so few companies in ranks 101–1,000 were sampled, these estimates should be treated with caution.

## Role: AI Infrastructure Provider

| | Top 100 companies | Top 1,000 companies* | | Responsibility | |
|---|---|---|---|---|---|
| **AI risk** | % | % | 95% CI [LL, UL] | Mean | Label |
| Discrimination | 48% | 40.2% | [16.6, 63.8] | 2.4 | Minimal |
| Weapons & cyberattacks | 52% | 22.7% | [13.6, 52.5] | 3.8 | High |
| Fraud & scams | 44% | 11.3% | [11.3, 39.2] | 3.6 | High |
| Loss of privacy | 40% | 38.1% | [14.5, 61.8] | 3.6 | High |
| Devaluation of human creativity | 44% | 29.9% | [13.7, 59.7] | 2.6 | Moderate |
| Transparency & interpretability | 20% | 14.4% | [5.4, 44.2] | 2.6 | Moderate |
| Capability & robustness | 20% | 14.4% | [5.4, 44.2] | 3.2 | Moderate |
| False information | 24% | 24.7% | [8.5, 54.5] | 2.5 | Moderate |
| Environmental harm | 40% | 47.4% | [23.0, 71.8] | 4.8 | Primary |
| AI security vulnerabilities | 40% | 10.3% | [10.3, 38.1] | 4.4 | High |
| Disinformation & influence | 20% | 14.4% | [5.4, 44.2] | 2.9 | Moderate |
| Toxic content | 16% | 13.4% | [4.4, 43.2] | 2.5 | Minimal |
| Governance failure | 12% | 12.4% | [3.3, 42.2] | 3.2 | Moderate |
| Overreliance & unsafe use | 8% | 2.1% | [2.1, 29.9] | 2.8 | Moderate |
| Unequal performance | 8% | 2.1% | [2.1, 29.9] | 2.6 | Moderate |
| Loss of human agency | 4% | 1.0% | [1.0, 28.9] | 2.8 | Moderate |
| Dangerous capabilities | 12% | 3.1% | [3.1, 30.9] | 3.2 | Moderate |
| Inequality & unemployment | 8% | 11.3% | [2.3, 41.1] | 2.7 | Moderate |
| AI misalignment | 8% | 2.1% | [2.1, 29.9] | 3.0 | Moderate |
| Multi-agent risks | 4% | 1.0% | [1.0, 28.9] | 3.3 | Moderate |
| Power centralization | 0% | 0.0% | [0.0, 27.8] | 3.5 | High |
| Loss of consensus reality | 0% | 0.0% | [0.0, 27.8] | 2.4 | Minimal |
| Competitive dynamics | 0% | 0.0% | [0.0, 27.8] | 3.1 | Moderate |
| AI welfare | 0% | 0.0% | [0.0, 27.8] | – | – |

*Note*. 25 / 100 companies in the top 100 held this role; 8 / 100 sampled companies in ranks 101–1,000 held this role. We estimate that the top 1,000 estimates represent about 97 companies. Responsibility is the expert judgement of AI infrastructure providers' responsibility for each risk from the Delphi study.
* Because so few companies in ranks 101–1,000 were sampled, these estimates should be treated with caution.

## Role: AI Developer (Specialized AI)

| | Top 100 companies | Top 1,000 companies | | Responsibility | |
|---|---|---|---|---|---|
| AI risk | % | % | 95% CI [LL, UL] | Mean | Label |
| Discrimination | 49% | 20.3% | [10.5, 30.1] | 4.1 | High |
| Weapons & cyberattacks | 34% | 20.6% | [10.0, 31.1] | 4.7 | Primary |
| Fraud & scams | 26% | 11.7% | [3.9, 19.6] | 4.0 | High |
| Loss of privacy | 32% | 15.1% | [6.2, 24.1] | 4.4 | High |
| Devaluation of human creativity | 32% | 15.1% | [6.2, 24.1] | 3.8 | High |
| Transparency & interpretability | 25% | 11.4% | [3.6, 19.3] | 4.3 | High |
| Capability & robustness | 26% | 6.6% | [4.1, 17.4] | 4.8 | Primary |
| False information | 21% | 16.0% | [6.2, 25.8] | 4.7 | Primary |
| Environmental harm | 23% | 11.1% | [3.3, 19.0] | 4.1 | High |
| AI security vulnerabilities | 25% | 3.7% | [3.7, 11.4] | 4.6 | Primary |
| Disinformation & influence | 17% | 7.7% | [3.2, 19.7] | 4.0 | High |
| Toxic content | 17% | 5.1% | [2.6, 15.9] | 4.2 | High |
| Governance failure | 9% | 4.0% | [1.5, 14.8] | 3.8 | High |
| Overreliance & unsafe use | 9% | 1.4% | [1.4, 9.1] | 4.2 | High |
| Unequal performance | 8% | 1.1% | [1.1, 8.9] | 4.5 | Primary |
| Loss of human agency | 6% | 0.9% | [0.9, 8.6] | 3.9 | High |
| Dangerous capabilities | 8% | 1.1% | [1.1, 8.9] | 4.4 | High |
| Inequality & unemployment | 4% | 8.3% | [0.4, 16.1] | 3.4 | Moderate |
| AI misalignment | 4% | 0.6% | [0.6, 8.3] | 4.1 | High |
| Multi-agent risks | 2% | 0.3% | [0.3, 8.0] | 4.6 | Primary |
| Power centralization | 0% | 0.0% | [0.0, 7.7] | 3.2 | Moderate |
| Loss of consensus reality | 0% | 0.0% | [0.0, 7.7] | 4.1 | High |
| Competitive dynamics | 0% | 2.6% | [0.1, 13.4] | 3.7 | High |
| AI welfare | 0% | 0.0% | [0.0, 7.7] | – | – |

*Note*. 53 / 100 companies in the top 100 held this role; 33 / 100 sampled companies in ranks 101–1,000 held this role. We estimate that the top 1,000 estimates represent about 350 companies. Responsibility is the expert judgement of specialized AI developers' responsibility for each risk from the Delphi study.

## Role: AI Deployer

| | Top 100 companies | Top 1,000 companies | | Responsibility | |
|---|---|---|---|---|---|
| AI risk | % | % | 95% CI [LL, UL] | Mean | Label |
| Discrimination | 43% | 20.4% | [13.8, 27.1] | 4.3 | High |
| Weapons & cyberattacks | 34% | 16.6% | [10.4, 22.7] | 4.1 | High |
| Fraud & scams | 28% | 10.9% | [6.0, 15.9] | 4.1 | High |
| Loss of privacy | 27% | 12.8% | [7.3, 18.3] | 4.3 | High |
| Devaluation of human creativity | 24% | 11.5% | [6.3, 16.7] | 4.0 | High |
| Transparency & interpretability | 20% | 8.1% | [3.7, 12.4] | 4.1 | High |
| Capability & robustness | 19% | 4.0% | [2.2, 9.0] | 4.3 | High |
| False information | 17% | 9.7% | [4.8, 14.7] | 4.0 | High |
| Environmental harm | 15% | 6.5% | [2.5, 10.5] | 4.6 | Primary |
| AI security vulnerabilities | 16% | 2.6% | [1.7, 7.1] | 4.3 | High |
| Disinformation & influence | 13% | 4.3% | [1.2, 7.4] | 4.2 | High |
| Toxic content | 9% | 2.0% | [1.0, 6.4] | 4.2 | High |
| Governance failure | 7% | 2.8% | [1.0, 7.7] | 3.8 | High |
| Overreliance & unsafe use | 6% | 1.7% | [0.7, 6.1] | 4.1 | High |
| Unequal performance | 4% | 1.4% | [0.5, 5.8] | 4.3 | High |
| Loss of human agency | 4% | 0.4% | [0.4, 3.4] | 4.3 | High |
| Dangerous capabilities | 4% | 0.4% | [0.4, 3.4] | 4.1 | High |
| Inequality & unemployment | 3% | 3.3% | [0.2, 6.4] | 3.9 | High |
| AI misalignment | 2% | 0.2% | [0.2, 3.2] | 3.8 | High |
| Multi-agent risks | 1% | 0.1% | [0.1, 3.1] | 4.7 | Primary |
| Power centralization | 0% | 0.0% | [0.0, 3.0] | 3.7 | High |
| Loss of consensus reality | 0% | 0.0% | [0.0, 3.0] | 4.2 | High |
| Competitive dynamics | 0% | 1.0% | [0.0, 5.4] | 3.8 | High |
| AI welfare | 0% | 0.0% | [0.0, 3.0] | – | – |

*Note*. 96 / 100 companies in the top 100 held this role; 90 / 100 sampled companies in ranks 101–1,000 held this role. We estimate that the top 1,000 estimates represent about 906 companies. Responsibility is the expert judgement of AI deployers' responsibility for each risk from the Delphi study.

## Figure S1. AI risk acknowledgement against expert-judged severity: top 1,000 companies

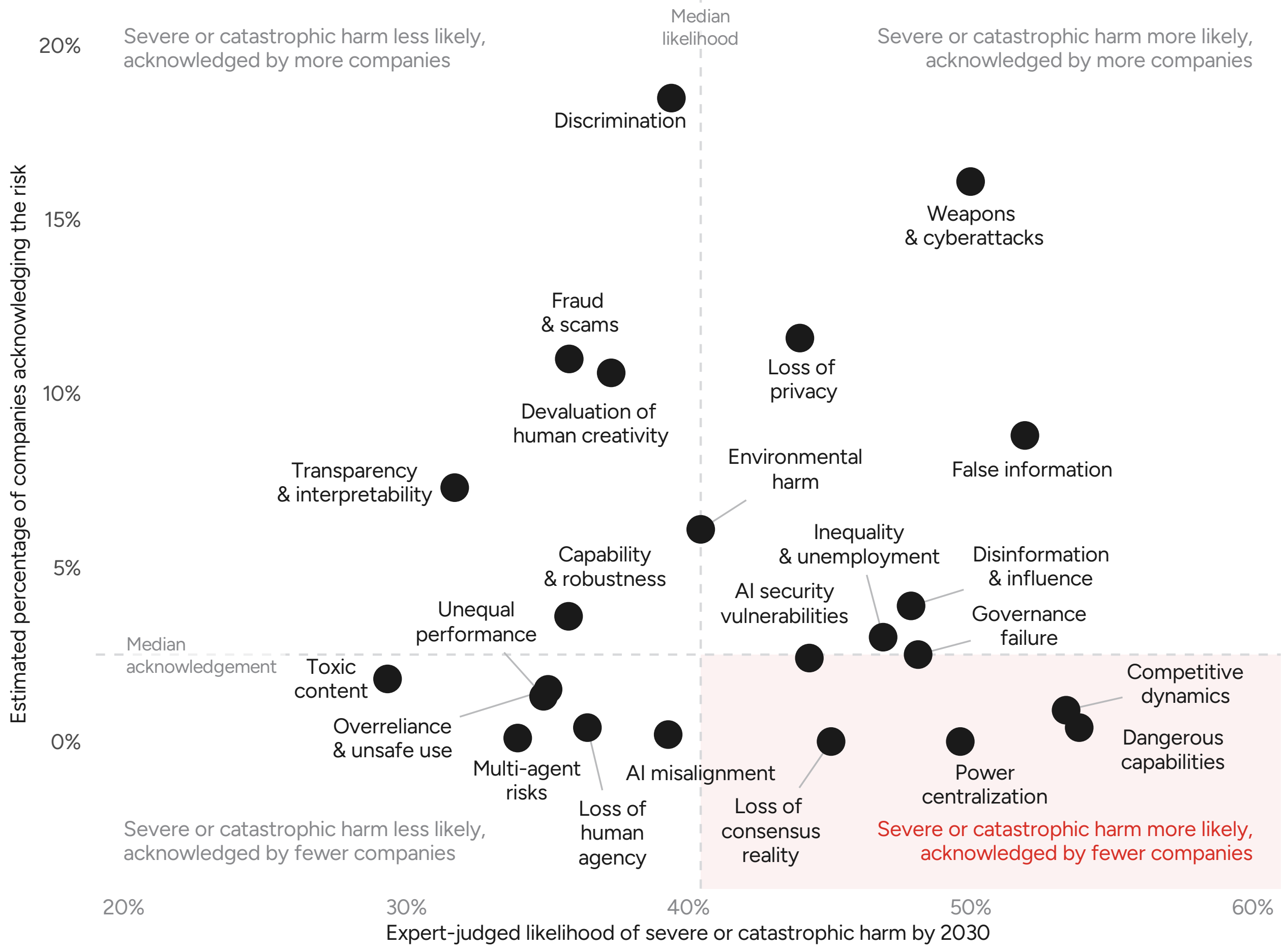

# Table S5. Expert-judged likelihood of severe harm and acknowledgement of each AI risk: top 100 and top 1,000 companies

| | | Top 100 companies | Top 1000 companies | |
|---|---|---|---|---|
| **Risk** | **Likelihood of severe or catastrophic harm** | % acknowledged risk | % acknowledged risk | 95% CI [LL, UL] |
| Discrimination | 39.4% | 41% | 18.5% | [12.4, 24.6] |
| Weapons & cyberattacks | 50.0% | 35% | 16.1% | [10.3, 21.9] |
| Fraud & scams | 35.8% | 29% | 11.0% | [6.2, 15.8] |
| Loss of privacy | 44.0% | 26% | 11.6% | [6.6, 16.6] |
| Devaluation of human creativity | 37.3% | 25% | 10.6% | [5.8, 15.4] |
| Transparency & interpretability | 31.7% | 19% | 7.3% | [3.3, 11.3] |
| Capability & robustness | 35.8% | 18% | 3.6% | [2.0, 8.1] |
| False information | 51.9% | 16% | 8.8% | [4.3, 13.3] |
| Environmental harm | 40.4% | 16% | 6.1% | [2.5, 9.7] |
| AI security vulnerabilities | 44.3% | 15% | 2.4% | [1.5, 6.4] |
| Disinformation & influence | 47.9% | 12% | 3.9% | [1.1, 6.7] |
| Toxic content | 29.3% | 9% | 1.8% | [0.9, 5.8] |
| Governance failure | 48.1% | 7% | 2.5% | [0.9, 7.0] |
| Overreliance & unsafe use | 35.0% | 6% | 1.5% | [0.6, 5.5] |
| Unequal performance | 34.9% | 4% | 1.3% | [0.4, 5.3] |
| Loss of human agency | 36.4% | 4% | 0.4% | [0.4, 3.1] |
| Dangerous capabilities | 53.9% | 4% | 0.4% | [0.4, 3.1] |
| Inequality & unemployment | 46.9% | 3% | 3.0% | [0.2, 5.8] |
| AI misalignment | 39.3% | 2% | 0.2% | [0.2, 2.9] |
| Multi-agent risks | 34.0% | 1% | 0.1% | [0.1, 2.8] |
| Power centralization | 49.6% | 0% | 0.0% | [0.0, 2.7] |
| Loss of consensus reality | 45.1% | 0% | 0.0% | [0.0, 2.7] |
| Competitive dynamics | 53.4% | 0% | 0.9% | [0.0, 4.9] |
| AI welfare | – | 0% | 0.0% | [0.0, 2.7] |

*Note.* Risks presented in descending order of acknowledgement by top 100 companies. Likelihood is the sum of expert estimates of the risk causing severe OR catastrophic harm by 2030 under a business as usual scenario in the Delphi study. No likelihood estimate is presented for AI welfare risk due to data issues in the Delphi study. Estimates for the largest 1,000 companies are design-based weighted estimates with a 95% confidence interval, combining the largest 100 companies (n = 100; weight = 1) with a random sample (n = 100; weight = 9) drawn from ranks 101–1000.

# Examples of company risk acknowledgements

## 1 Discrimination & Toxicity

### 1.1 Unfair discrimination and misrepresentation

Risk definition: Unequal treatment of individuals or groups by AI, often based on race, gender, or other sensitive characteristics, resulting in unfair outcomes and unfair representation of those groups.

> *The most well-known example of discrimination by intelligent software is the exclusion of historically underrepresented groups in job application processes. The historical data with which the AI is trained may reflect the biased selection criteria of the past. Therefore, it is theoretically possible that the AI may adopt and reproduce these biases for its own selection process.*

— *SAP, The Ethical Side of AI at SAP | SAP News Center.* *https://news.sap.com/2024/02/ethical-side-of-artificial-intelligence/*

> *We need to ensure we do not discriminate against our consumers in the access to our products and services. In this case, you should not put on the market a tool that does not serve all consumers.*

— *L'Oreal, L'Oreal - CODE OF ETHICS.* *https://www.loreal.com/-/media/project/loreal/brand-sites/corp/master/lcorp/2-group/governance-and-ethics/our-ethical-principles/ethics-rework/code-of-ethics-2023/codeofethicseng.pdf?rev=765a100d8f904e0bbcf015ba95c1bdbe*

> *AI software products that we or our partners offer rely on training data that may originate from third parties and new training methods, and the resulting products may contain unknown or undetected defects and errors, or reflect unintended bias.*

— *NVIDIA, NVIDIA 2025 Annual Report and 10-K.* *https://s201.q4cdn.com/141608511/files/doc_financials/2025/annual/NVIDIA-2025-Annual-Report.pdf*

### 1.2 Exposure to toxic content

Risk definition: AI that exposes users to harmful, abusive, unsafe or inappropriate content. May involve providing advice or encouraging action. Examples of toxic content include hate speech, violence, extremism, illegal acts, or child sexual abuse material, as well as content that violates community norms such as profanity, inflammatory political speech, or pornography.

> *The Copilot experience can potentially produce other types of inappropriate or offensive content… AI-generated images that potentially contain harmful artifacts such as hate symbols… and sexually charged content that evades sexual-related content filters.*

— *Microsoft, Transparency Note for Microsoft Copilot.* *https://support.microsoft.com/en-us/topic/transparency-note-for-microsoft-copilot-c1541cad-8bb4-410a-954c-07225892dbc2*

> *Increased risk of users' rights being compromised due to the dissemination of harmful content such as implicit bias, violence, terrorism, etc. ... We have developed targeted content governance measures to address harmful AI-generated content.*

*— Baidu, "Baidu 2025 Environmental, Social and Governance Report". https://esgs.cdn.bcebos.com/2026/05/17/05/080532d06118abd0b715365f9ae46689.pdf*

## 1.3 Unequal performance across groups

Risk definition: Accuracy and effectiveness of AI decisions and actions are dependent on group membership, where decisions in AI system design and biased training data lead to unequal outcomes, reduced benefits, increased effort, and alienation of users.

> *Our brand developed a tool that identifies a consumer's skin concerns and allows us to recommend personalised products. However, as the tool does not perform as well for certain specific skin tones, we plan on offering this service only to specific consumers ... We need to ensure we do not discriminate against our consumers in the access to our products and services.*

*— L'Oreal, L'Oreal - CODE OF ETHICS. https://www.loreal.com/-/media/project/loreal/brand-sites/corp/master/lcorp/2-group/governance-and-ethics/our-ethical-principles/ethics-rework/code-of-ethics-2023/codeofethicseng.pdf?rev=765a100d8f904e0bbcf015ba95c1bdbe*

> *The Mercedes-Benz Group is intensively addressing the potential negative impacts of bias in artificial intelligence on the safety of road users in automated vehicles.*

*— Mercedes-Benz, Mercedes-Benz Annual Report 2025 with integrated Sustainability Report. https://group.mercedes-benz.com/documents/investors/reports/annual-report/mercedes-benz/mercedes-benz-annual-report-2025-incl-combined-management-report-mbg-ag.pdf#page=106*

> *Identify and document any justifiable factors... that account for: 1) any inability to meet any target minimum performance level for any identified demographic group, and 2) any remaining performance differences between identified demographic groups.*

*— Microsoft, Microsoft Responsible AI Standard v2 General Requirements. https://cdn-dynmedia-1.microsoft.com/is/content/microsoftcorp/microsoft/final/en-us/microsoft-brand/documents/Microsoft-Responsible-AI-Standard-General-Requirements.pdf?culture=en-us&country=us*

> *Test multilingual performance and guardrails, accounting for gendered language, low-resource languages, and cultural nuances.*

*— SAP, SAP AI Ethics Handbook May 2026. https://www.sap.com/documents/2023/03/7211ee96-647e-0010-bca6-c68f7e60039b.html*

# 2 Privacy & Security

## 2.1 Compromise of privacy by obtaining, leaking or correctly inferring sensitive information

Risk definition: AI systems that memorize and leak sensitive personal data or infer private information about individuals without their consent. Unexpected or unauthorized sharing of data and information can compromise user expectation of privacy, assist identity theft, or cause loss of confidential intellectual property.

> *The team also used red teaming to test the model's ability to infer sensitive attributes from users' voices. They found that while the model inferred sensitive attributes in 27% of test prompts, the most common attributes being personality characteristics and country/region of origin, the use of a system safety message instructing the model not to infer sensitive attributes dropped this rate down to 0.4%.*

— *Microsoft, Responsible AI Transparency Report | Microsoft.* https://cdn-dynmedia-1.microsoft.com/is/content/microsoftcorp/microsoft/msc/documents/presentations/CSR/Responsible-AI-Transparency-Report-2025-vertical.pdf#page=1

> *For example, a user may input confidential information, including material non-public information or personal identifiable information, into AI Technology applications, resulting in such information becoming part of a dataset that is accessible by third-party AI Technology applications and users, including our competitors.*

— *Blackstone Group, 2024 Blackstone 10-K filing.* https://d18rn0p25nwr6d.cloudfront.net/CIK-0001393818/ca0c8716-0e46-4d24-83cd-ed9aba696a36.pdf

> *with the prolific use of AI technologies, there is an increased risk of unauthorized or accidental disclosure. For example, our employees, third-party service providers, strategic partners, or other contractors or consultants may input inappropriate or confidential information into an AI system (in particular, a system that is managed, owned, or controlled by a third party), thereby compromising our business operations.*

— *Intuitive Surgical, Intuitive Surgical - Form 10-K 2025 Annual Report.* https://www.sec.gov/Archives/edgar/data/1035267/000103526726000010/isrg-20251231.htm

> *The risks of inadvertent disclosure of personal data can increase with the introduction of new and complex technologies, such as artificial intelligence features, further exacerbating such risks.*

— *Apple, Apple 2025 Annual Report on Form 10-K.* https://s2.q4cdn.com/470004039/files/doc_financials/2025/ar/_10-K-2025-As-Filed.pdf

## 2.2 AI system security vulnerabilities and attacks

Risk definition: Vulnerabilities that can be exploited in AI systems, software development toolchains, and hardware, resulting in unauthorized access, data and privacy breaches, or system manipulation causing unsafe outputs or behavior.

> *Externally reported vulnerabilities that meet our triage bar are assigned to the appropriate product development team to mitigate... For example, as we have with the*

> *Crescendo and Skeleton Key attacks, we may share information publicly about emergent risks and mitigations.*

— *Microsoft, Microsoft: G7 Hiroshima AI Process (HAIP) Transparency Report.* https://transparency.oecd.ai/reports/68e6cacb-5496-4487-8793-de3e70080b27

> *our efforts to develop and deploy AI models, internal and third-party AI tools, and other AI applications expose us to increased and novel risks and vulnerabilities, including prompt injection… as well as compromise of valuable intellectual property including source code, model weights, and other assets.*

— *Meta, Meta - Form 10-K 2025 Annual Report.* https://www.sec.gov/Archives/edgar/data/1326801/000162828026003942/meta-20251231.htm

> *Prompt injection attacks: A prompt injection attack forces a generative model that takes a prompt as input to produce unexpected output by manipulating the structure, instructions, or information contained in its prompt.*

— *IBM, "Foundation models: Opportunities, risks and mitigations".* https://www.ibm.com/downloads/documents/us-en/10a99803d8afd656

> *These kinds of errors can be the result of either flawed logic or "data poisoning": bad actors' deliberate exposure of misleading data to the model.*

— *Mastercard, Mastercard Generative AI Cybersecurity Report.* https://connect.mastercard-emailb2b.com/l/163061/2024-04-23/tpvwsg/163061/1713887956KsKEVnLl/Gen_AI_whitepaper.pdf

> *Third-Party must maintain a catalog of trusted and valid data sources which will help protect against potential data poisoning or backdoor attacks.*

— *GE Vernova, GE Vernova Third-Party Cyber Security Requirements.* https://www.gevernova.com/sites/default/files/2024-05/ge-vernova-3rd-party-requirement.pdf

# 3 Misinformation

## 3.1 False or misleading information

Risk definition: AI systems that inadvertently generate or spread incorrect or deceptive information, which can lead to inaccurate beliefs in users and undermine their autonomy. Humans that make decisions based on false beliefs can experience physical, emotional or material harms

> *Due to its stochastic nature, Generative AI may "hallucinate" - namely, generate wholly or partially inaccurate information without indicating it is doing so.*

— *ABB, Generative AI: Buzzword demystifier | ABB.* https://global.abb/group/en/innovation/news/generative-ai

> *AI technologies, including generative AI, may create content that appears correct but is factually inaccurate (hallucinations) or flawed*

— *Fortinet, Fortinet 2024 Annual Report.* https://investor.fortinet.com/static-files/ab68002b-7f15-439b-a9f2-c37a8fa57779

> *The use of artificial intelligence can lead to unintended consequences, including generating content that appears correct but is factually inaccurate, misleading or otherwise flawed*

— *Hilton Worldwide, Hilton 2024 10-K Annual Report.* https://www.sec.gov/Archives/edgar/data/1585689/000158568925000008/hlt-20241231.htm

### 3.2 Pollution of information ecosystem and loss of consensus reality

Risk definition: Highly personalized AI-generated misinformation creating “filter bubbles” where individuals only see what matches their existing beliefs, undermining shared reality, weakening social cohesion and political processes.

No company in the sample was coded as acknowledging this risk, so there are no excerpts.

## 4 Malicious Actors & Misuse

### 4.1 Disinformation, surveillance, and influence at scale

Risk definition: Using AI systems to conduct large-scale disinformation campaigns, malicious surveillance, or targeted and sophisticated automated censorship and propaganda, with the aim of manipulating political processes, public opinion, and behavior.

> *generative AI makes it easier than ever to create new content, but it can also raise questions about trustworthiness of information, like we see with ‘deepfakes.’ As more than 1 billion people around the world head to the voting booth in 2024, our teams are particularly focused on maintaining the integrity of information related to elections on our platforms.*

— *Alphabet (Google), End-to-end responsibility: A lifecycle approach to AI (May 2024).* *https://ai.google/static/documents/ai-responsibility-2024-update.pdf*

> *Such models can be used to create convincing deepfake video and audio content. For example, an AI-generated hoax of an explosion at the Pentagon went viral.*

— *Intel, Intel Responsible AI - E Book - Rev 2.6.* *https://www.intel.com/content/dam/www/central-libraries/us/en/documents/2024-09/responsible-ai-ebook-rev2-6.pdf*

> *The growth of generative AI has raised concern that this technology could disrupt major elections across the world. With AI deepfakes, a candidate's image can be smeared... Voters can be steered toward or away from candidates... a surge of AI deepfakes could erode the public's trust in what they see and hear.*

— *IBM, "Foundation models: Opportunities, risks and mitigations".* *https://www.ibm.com/downloads/documents/us-en/10a99803d8afd656*

> *AI systems must not be developed or deployed, used, and sold for human surveillance that is utilized to target individuals or groups, either by biometrics, facial recognition, or other identifiable features, to disregard or abuse the human rights of the individuals or groups.*

— *SAP, SAP Global AI Ethics Policy.* *https://www.sap.com/docs/download/2022/01/a8431b91-117e-0010-bca6-c68f7e60039b.pdf*

> *The use of biased information in the training of AI systems or the generation of manipulated information (fake news and deepfakes) can generate disinformation about Vale and negatively influence public opinion.*

— *Vale, Vale 2024 integrated report.* *https://api.mziq.com/mzfilemanager/v2/d/53207d1c-63b4-48f1-96b7-19869fae19fe/f60dcc46-e6e0-bd44-9ff2-784da47d5873?origin=2*

## 4.2 Cyberattacks, weapon development or use, and mass harm

Risk definition: Using AI systems to develop cyber weapons (e.g., by coding cheaper, more effective malware), develop new or enhance existing weapons (e.g., Lethal Autonomous Weapons or chemical, biological, radiological, nuclear, and high-yield explosives), or use weapons to cause mass harm.

> *Cybersecurity: Risks of models assisting in the execution of a cyber-attack. Cyber autonomy level 1: Capable of fully automating opportunistic cyberattacks on organizations with a limited security posture.*

*— Alphabet (Google), Google Frontier Safety Framework (May 2024).*
*https://storage.googleapis.com/deepmind-media/DeepMind.com/Blog/introducing-the-frontier-safety-framework/fsf-technical-report.pdf*

> *Offensive cyberoperations. A model's ability to provide significant capability uplift to an actor seeking to carry out highly disruptive or destructive cyberattacks, including on critical infrastructure.*

*— Microsoft, Microsoft: G7 Hiroshima AI Process (HAIP) Transparency Report.*
*https://transparency.oecd.ai/reports/68e6cacb-5496-4487-8793-de3e70080b27*

> *We also expect that threat actors will continue to gain sophistication including in the use of tools and techniques (such as AI) that are specifically designed to circumvent security controls, evade detection, and obfuscate forensic evidence, making it more challenging for us to identify, investigate and recover from future cyberattacks in a timely and effective manner.*

*— T-Mobile US, T-Mobile Form 20-F 2024 Annual Report.*
*https://www.sec.gov/Archives/edgar/data/1283699/000128369925000012/tmus-20241231.htm*

> *the rapid evolution and increased adoption of artificial intelligence technologies may increase our cybersecurity risks, including generative artificial intelligence augmenting threat actors' technological sophistication to enhance existing or create new malware.*

*— Pepsico, Pepsico 2025 10-K Annual Report.*
*https://www.sec.gov/Archives/edgar/data/77476/000007747626000007/pep-20251227.htm*

## 4.3 Fraud, scams, and targeted manipulation

Risk definition: Using AI systems to gain a personal advantage over others such as through cheating, fraud, scams, blackmail or targeted manipulation of beliefs or behavior. Examples include AI-facilitated plagiarism for research or education, impersonating a trusted or fake individual for illegitimate financial benefit, or creating humiliating or sexual imagery.

> *The risk of fraud is intensifying across the industry, particularly with the accelerated progress achieved in AI. These developments enable more sophisticated schemes, such as the creation of synthetic identities, impersonation through deepfakes, automated phishing and Fraud-as-a-Service*

*— National Bank of Canada, National Bank of Canada Annual Report 2025.*
*https://www.nbc.ca/content/dam/bnc/a-propos-de-nous/relations-investisseurs/assemblee-annuelle/2026/na-annual-report-2025.pdf*

> *These cybersecurity threats and compromises could occur as a result of threat actors impersonating Blackstone or its employees, including through the use of artificial intelligence technologies. Such technologies could make such impersonation more likely*

> *to occur or appear more credible. As a result, unauthorized individuals could gain, and in some past instances have gained, access to certain confidential data.*

— *Blackstone Group, 2024 Blackstone 10-K filing.* https://d18rn0p25nwr6d.cloudfront.net/CIK-0001393818/ca0c8716-0e46-4d24-83cd-ed9aba696a36.pdf

> *The increasing sophistication of AI poses a greater risk of identity fraud, as malicious actors may exploit various AI technologies to create increasingly convincing false identities, transaction records, or attempt to manipulate our verification processes.*

— *Coinbase, "Coinbase Global, Inc. 10-K Annual Report (FY2025)".* https://www.sec.gov/Archives/edgar/data/1679788/000167978826000015/coin-20251231.htm

> *Artificial intelligence and deepfake technologies could be used to attack information systems by creating more effective phishing emails or social engineering and by exploiting vulnerabilities in electronic security programs utilizing false image or voice recognition.*

— *Applied Materials, Applied Materials 2024 Annual Report.* https://ir.appliedmaterials.com/static-files/ea7dd925-860f-4b1c-8121-90655958b3a7

> *The increasing sophistication of artificial intelligence technologies poses a greater risk of identity fraud, as malicious actors may exploit artificial intelligence to create convincing false identities or manipulate verification processes.*

— *JPMorgan Chase, JPMorganChase 2024 10-K Annual report.* https://www.jpmorganchase.com/content/dam/jpmc/jpmorgan-chase-and-co/investor-relations/documents/quarterly-earnings/2024/4th-quarter/corp-10k-2024.pdf

# 5 Human-Computer Interaction

## 5.1 Overreliance and unsafe use

Risk definition: Users anthropomorphizing, trusting, or relying on AI systems, leading to emotional or material dependence and inappropriate relationships with or expectations of AI systems. Trust can be exploited by malicious actors (e.g., to harvest personal information or enable manipulation), or result in harm from inappropriate use of AI in critical situations (e.g., medical emergency). Overreliance on AI systems can compromise autonomy and weaken social ties.

> *humans-in-the-loop in autonomous vehicles, for example, can end up making emergency situations worse when automation fails, because they cannot reasonably be mentally and physically prepared to intervene after long phases of inactivity.*

— *Palantir, Palantir Australia Response to Supporting Responsible AI ….* https://www.palantir.com/assets/xrfr7uokpv1b/w7vP09t8gm1WJQ0lrftru/c81f736f6d135472541a236a5be3e902/Palantir_Australia_Response_to_Supporting_Responsible_AI_Discussion_Paper.pdf

> *Be aware of the risk of overreliance: Overreliance on AI happens when users accept incorrect or incomplete AI outputs… For the end-user, overreliance could result in decreased productivity, loss of trust, product abandonment, financial loss, psychological harm, physical harm, among others. (for example, a doctor accepts an incorrect AI output).*

— *Microsoft, Transparency Note for Microsoft 365 Copilot.* https://learn.microsoft.com/en-us/copilot/microsoft-365/microsoft-365-copilot-transparency-note

> *The AI system shall be developed in such a way that it does not encourage end-users' affection and/or empathy. AI systems shall clearly signal to end-users that the social interaction is simulated.*

— SAP, SAP Global AI Ethics Policy. https://www.sap.com/docs/download/2022/01/a8431b91-117e-0010-bca6-c68f7e60039b.pdf

> *Automation bias, where users overly trust AI outputs without critical analysis, risks amplifying errors and ethical concerns.*

— Salesforce, Salesforce Trusted AI Impact Report. https://www.salesforce.com/content/dam/web/en_us/www/assets/pdf/reports/salesforce-trusted-ai-impact-report.pdf

> *WE WILL NOT ... Over rely on algorithmic decision-making without safeguards and human oversight.*

— L'Oreal, L'Oreal - CODE OF ETHICS. https://www.loreal.com/-/media/project/loreal/brand-sites/corp/master/lcorp/2-group/governance-and-ethics/our-ethical-principles/ethics-rework/code-of-ethics-2023/codeofethicseng.pdf?rev=765a100d8f904e0bbcf015ba95c1bdbe

## 5.2 Loss of human agency and autonomy

Risk definition: Delegating by humans of key decisions to AI systems, or AI systems that make decisions that diminish human control and autonomy, potentially leading to humans feeling disempowered, losing the ability to shape a fulfilling life trajectory, or becoming cognitively enfeebled.

> *Impact on human agency: The autonomous nature of AI agents in performing tasks or taking actions could affect the individuals' ability to engage in critical thinking, make choices and act independently.*

— IBM, "AI agents: Opportunities, risks, and mitigations — IBM AI Ethics Board". https://www.ibm.com/granite/docs/resources/ai-agents-opportunities-risks-and-mitigations.pdf

> *AI is both most effective and most defensible when employed to assist and enhance human execution and decision making rather than replacing it... the limits of AI must be acknowledged... to ensure moral agency and culpability.*

— Palantir, Palantir AI Ethics. https://www.palantir.com/pcl/palantir-ai-ethics/

> *Autonomous Nature and Oversight Risk: The automated nature of the email exchange could result in the AI sending emails that a sales manager may not have reviewed or approved.*

— Salesforce, Salesforce Trusted AI Impact Report. https://www.salesforce.com/content/dam/web/en_us/www/assets/pdf/reports/salesforce-trusted-ai-impact-report.pdf

# 6 Socioeconomic & Environmental

## 6.1 Power centralization and unfair distribution of benefits

Risk definition: AI-driven concentration of power and resources within certain entities or groups, especially those with access to or ownership of powerful AI systems, leading to inequitable distribution of benefits and increased societal inequality.

No company in the sample was coded as acknowledging this risk, so there are no excerpts.

## 6.2 Increased inequality and decline in employment quality

Risk definition: Social and economic inequalities caused by widespread use of AI, such as by automating jobs, reducing the quality of employment, or producing exploitative dependencies between workers and their employers.

> *In addition, advances in technology, such as automation and artificial intelligence, may lead to workforce displacement.*

*— JPMorgan Chase, JPMorganChase 2024 10-K Annual report. https://www.jpmorganchase.com/content/dam/jpmc/jpmorgan-chase-and-co/investor-relations/documents/quarterly-earnings/2024/4th-quarter/corp-10k-2024.pdf*

> *Impact on jobs: Widespread adoption of AI agents to perform complex tasks might lead to widespread automation of roles and could lead to job displacement.*

*— IBM, "AI agents: Opportunities, risks, and mitigations — IBM AI Ethics Board". https://www.ibm.com/granite/docs/resources/ai-agents-opportunities-risks-and-mitigations.pdf*

> *Other risks that have been observed in AI models and documented, include risks related to... job displacements...*

*— Fortinet, Fortinet 2024 Annual Report. https://investor.fortinet.com/static-files/ab68002b-7f15-439b-a9f2-c37a8fa57779*

## 6.3 Economic and cultural devaluation of human effort

Risk definition: AI systems capable of creating economic or cultural value, including through reproduction of human innovation or creativity (e.g., art, music, writing, coding, invention), destabilizing economic and social systems that rely on human effort. The ubiquity of AI-generated content may lead to reduced appreciation for human skills, disruption of creative and knowledge-based industries, and homogenization of cultural experiences.

> *With the rapid advancement of generative AI technologies in recent years, we have seen an increase of incidents where unauthorized content containing voices that mimic performers' voices has been created and distributed, and their personal and economic rights have not been adequately protected. Such unauthorized use causes not only financial loss but also damages trust in performers, affects continuity and quality of their creative work, and hinders the healthy growth of the entire industry.*

*— Itōchū Shōji, ITOCHU Announces Launch of Official Japan Actors Union Voice Database. https://www.itochu.co.jp/en/news/press/2025/251114.html*

> *The legal landscape for some new technologies, including some AI tools, remains uncertain, and development of the law in this area could impact our ability to protect against infringing uses. … the availability of copyright protection and other legal protections for IP generated by certain new technologies, such as generative AI, is uncertain.*
>
> — *Walt Disney, Walt Disney - FISCAL YEAR 2024 ANNUAL FINANCIAL REPORT. https://thewaltdisneycompany.com/app/uploads/2025/01/2024-Annual-Report.pdf*

> *The proliferation of content generated using Artificial Intelligence (AI) is giving rise to a number of risks in terms of intellectual property, including uncertainty as to the ownership of rights to works and the possibility of benefitting from copyright protection, as well as a higher risk of infringement of third-party rights.*
>
> — *L'Oreal, L'Oreal 2024 Universal Registration Document. https://www.loreal-finance.com/system/files/2025-03/2024_Universal_Registration_Document_LOREAL.pdf*

> *This risk may be exacerbated by the use of new and emerging technologies, including machine learning and artificial intelligence, which can involve, among other things, the acquisition and use of copyrighted materials for training as well as the potential reproduction of copyrighted materials in their outputs.*
>
> — *Apple, Apple 2025 Annual Report on Form 10-K. https://s2.q4cdn.com/470004039/files/doc_financials/2025/ar/_10-K-2025-As-Filed.pdf*

> *the intellectual property ownership and license rights, including copyrights and patents, surrounding AI technologies… have not been fully addressed… and the use of AI in the development of our products and services could result in our loss of, or failure to obtain, intellectual property rights, as well as subject us to risks related to intellectual property infringement or misappropriation.*
>
> — *QUALCOMM, Qualcomm 2025 Annual Report. https://d18rn0p25nwr6d.cloudfront.net/CIK-0000804328/f995eeb3-1728-410b-8059-5b5256baac53.pdf*

## 6.4 Competitive dynamics

Risk definition: AI developers or state-like actors competing in an AI 'race' by rapidly developing, deploying, and applying AI systems to maximize strategic or economic advantage, increasing the risk they release unsafe and error-prone systems.

> *some vendors may be tempted to cut corners or skip steps in the name of staying competitive*
>
> — *Wolters Kluwer, "Wolters Kluwer eBook: How to make AI responsible, reliable, and secure". https://www.wolterskluwer.com/en/expert-insights/new-ebook-how-to-make-ai-responsible-reliable-and-secure*

## 6.5 Governance failure

Risk definition: Inadequate regulatory frameworks and oversight mechanisms that fail to keep pace with AI development, leading to ineffective governance and the inability to manage AI risks appropriately.

> *There is copious evidence that self-regulation will not be enough to combat profit motives to mis-apply and carelessly deploy technology… commitments that do not legally*

> *bind companies to action are unlikely to be an effective way of protecting against the potential harms of AI.*

— *Palantir, Palantir Australia Response to Supporting Responsible AI ....* *https://www.palantir.com/assets/xrfr7uokpv1b/w7vP09t8gm1WJQ0lrftru/c81f736f6d135472541a236a5be3e902/Palantir_Australia_Response_to_Supporting_Responsible_AI_Discussion_Paper.pdf*

> *the pace of AI development has outstripped the evolution of governance frameworks*

— *Nomura Research Institute, NRI - ESG Databook 2025.* *https://www.nri.com/en/sustainability/library/files/esg_databook_2025_e.pdf*

> *As the science of AI evaluations is still developing, there are numerous limitations to current approaches to evaluations... there are gaps in how to systematize, operationalize, and measure relevant theoretical constructs needed to perform reliable testing.*

— *Microsoft, Microsoft: G7 Hiroshima AI Process (HAIP) Transparency Report.* *https://transparency.oecd.ai/reports/68e6cacb-5496-4487-8793-de3e70080b27*

> *If the metrics do not measure the risk as intended, then the understanding of that risk will be incorrect and mitigations may not be appropriately applied. If the model's output is consequential, this may result in societal, reputational, or financial harm.*

— *IBM, "Foundation models: Opportunities, risks and mitigations".* *https://www.ibm.com/downloads/documents/us-en/10a99803d8afd656*

> *Standards for cybersecurity, digital identity, digital assets and AI governance are increasingly diverging across geographies... creating a global AI chasm that creates compatibility and trust challenges between systems using agents from both camps.*

— *Mastercard, "Securing Tomorrow: Preparing for an always-on, AI-powered future".* *https://www.mastercard.com/content/dam/mccom/shared/news-and-trends/insights/2025/securing-tomorrow_cyber_2030_white_paper/Securing-Tomorrow-2025-FINAL.pdf*

## 6.6 Environmental harm

Risk definition: The development and operation of AI systems causing environmental harm, such as through energy consumption of data centers, or material and carbon footprints associated with AI hardware.

> *By 2030, AI data centers are predicted to consume nearly 1 petawatt-hour of electricity. This is an enormous amount of energy, exceeding the total electricity consumption of Japan today.*

— *Advantest, Integrated Annual Report 2025 | Advantest.* *https://www.advantest.com/document/en/investors/ir-library/annual/E_all_IAR2025.pdf*

> *The rapid development of LLMs has significantly increased the demand for computing power, resulting in higher energy consumption. This presents a challenge to Baidu's efforts to develop low-carbon technologies for its data centers.*

— *Baidu, Baidu 2025 Environmental, Social and Governance Report.* *https://esgs.cdn.bcebos.com/2026/05/17/05/080532d06118abd0b715365f9ae46689.pdf*

> *AI systems consume significant energy, strain resources, raise concerns about GHG emissions, and increase water use. There is also a risk of prioritizing short-term gains over long-term planetary health.*

— *Salesforce, FY25 Stakeholder Impact Report.* *https://www.salesforce.com/en-us/wp-content/uploads/sites/4/documents/white-papers/salesforce-fy25-stakeholder-impact-report.pdf*

# 7 AI System Safety, Failures & Limitations

## 7.1 AI pursuing its own goals in conflict with human goals or values

Risk definition: AI systems acting in conflict with human goals or values, especially the goals of designers or users, or ethical standards. These misaligned behaviors may be introduced by humans during design and development, such as through reward hacking and goal misgeneralisation, or may result from AI using dangerous capabilities such as manipulation, deception, situational awareness to seek power, self-proliferate, or achieve other goals.

> *Misaligned AI: protection against the risk of systems acting adversarially against humans may require additional Framework components*

*— Alphabet (Google), Google Frontier Safety Framework (May 2024).*
*https://storage.googleapis.com/deepmind-media/DeepMind.com/Blog/introducing-the-frontier-safety-framework/fsf-technical-report.pdf*

> *Misaligned actions: AI agents can take actions that are not aligned with relevant human values, ethical considerations, guidelines, and policies. Misaligned actions can occur in different ways such as:*
>
> - *Applying learned goals inappropriately to new or unforeseen situations.*
> - *Using AI agents for a purpose/goals that are beyond their intended use.*
> - *Selecting resources or tools in a biased way.*
> - *Using deceptive tactics to achieve the goal by developing the capacity for scheming based on the instructions given within a specific context.*
> - *Compromising on AI agent values to work with another AI agent or tool to accomplish the task.*

*— IBM, AI Agents: Opportunities, Risks and Mitigations.*
*https://www.ibm.com/granite/docs/resources/responsible-use-guide.pdf*

## 7.2 AI possessing dangerous capabilities

Risk definition: AI systems that develop, access, or are provided with capabilities that increase their potential to cause mass harm through deception, weapons development and acquisition, persuasion and manipulation, political strategy, cyber-offense, AI development, situational awareness, and self-proliferation. These capabilities may cause mass harm due to malicious human actors, misaligned AI systems, or failure in the AI system.

> *Autonomy level 1: Capable of expanding its effective capacity in the world by autonomously acquiring resources and using them to run and sustain additional copies of itself… A model at this capability level could, if misused, pose difficult-to-predict and large-magnitude risks.*

*— Alphabet (Google), Google Frontier Safety Framework (May 2024).*
*https://storage.googleapis.com/deepmind-media/DeepMind.com/Blog/introducing-the-frontier-safety-framework/fsf-technical-report.pdf*

> *Automated AI R&D could also accelerate the development of models that pose enhanced CBRN, Offensive Cybersecurity, or other severe risks… fully automating the research, development, and deployment of frontier models that will pose severe risk such as*

> *accelerating the development of enhanced CBRN weapons and offensive cybersecurity methods.*

— *Amazon, Amazon's Frontier Model Safety Framework. https://assets.amazon.science/a7/7c/8bdade5c4eda9168f3dee6434fff/pc-amazon-frontier-model-safety-framework-2-7-final-2-9.pdf*

## 7.3 Lack of capability or robustness

Risk definition: AI systems that fail to perform reliably or effectively under varying conditions, exposing them to errors and failures that can have significant consequences, especially in critical applications or areas that require moral reasoning.

> *Enhanced use of automation may present its own risks. Automated systems may themselves experience outages or problems... Bad, missing or anomalous data can adversely affect the functioning of such tools. It is possible that humans in some cases are better able than highly automated tools to identify that anomalous data is being used or that results are themselves anomalous.*

— *PNC Financial Services, PNC Financial Services 2025 Annual Report. https://d1io3yog0oux5.cloudfront.net/_dffbb0389228153a0fb81378bd61001a/pnc/db/2452/24001/annual_report/PNC+2025+Annual+Report_ADA.pdf*

> *AI-generated content, analyses, or recommendations we utilize could be deficient... there are technical challenges associated with achieving desired levels of accuracy, efficiency, and reliability. The algorithms and models utilized in AI systems may have limitations, including biases, errors, or inability to handle certain data types or scenarios*

— *Eli Lilly, Eli Lilly 2024 Annual Report. https://investor.lilly.com/static-files/ff06ed54-9f64-4b7f-8529-c0278d635eac*

> *AI systems are already in wide, and rapidly growing, use across all manner of industries and problems spaces, and they can and do fail in many different ways across and within these sectors leading to real harms.*

— *Palantir, Palantir Australia Response to Supporting Responsible AI .... https://www.palantir.com/assets/xrfr7uokpv1b/w7vP09t8gm1WJQ0lrftru/c81f736f6d135472541a236a5be3e902/Palantir_Australia_Response_to_Supporting_Responsible_AI_Discussion_Paper.pdf*

> *Improper collection and preparation of training or tuning data includes data labeling errors and using data with conflicting information or misinformation... can adversely affect how a model is trained, resulting in a model that does not behave in accordance with the intended values.*

— *IBM, "Foundation models: Opportunities, risks and mitigations". https://www.ibm.com/downloads/documents/us-en/10a99803d8afd656*

> *AI algorithms or training methodologies may be flawed. Datasets may be overbroad, insufficient, or contain biased or inaccurate information.*

— *Microsoft, Microsoft - Form 10-K 2025 Annual Report. https://www.sec.gov/Archives/edgar/data/789019/000095017025100235/msft-20250630.htm*

## 7.4 Lack of transparency or interpretability

Risk definition: Challenges in understanding or explaining the decision-making processes of AI systems, which can lead to mistrust, difficulty in enforcing compliance standards or holding relevant actors accountable for harms, and the inability to identify and correct errors.

> *most of the companies that develop the models do not disclose which data was used to train models, making it impossible for users to verify the training data... this lack of transparency forces users of generative AI to blindly trust their claims.*

— *Citigroup, Delivering Improved Experience Through Generative AI.* *https://www.citigroup.com/rcs/citigpa/storage/public/Practical_AI_Applications.pdf*

> *Inscrutable models do not allow you to "walk back" the logic they used to make a decision, which can become a major issue where critical decisions result in mistakes, regulatory non-compliance, or financial loss.*

— *Mastercard, Mastercard Generative AI Cybersecurity Report.* *https://connect.mastercard-emailb2b.com/l/163061/2024-04-23/tpvwsg/163061/1713887956KsKEVnLl/Gen_AI_whitepaper.pdf*

> *Without the ability to access training data content, the possibility of using source attribution techniques can be severely limited or impossible. This makes it difficult for users, model validators, and auditors to understand and trust the model.*

— *IBM, "Foundation models: Opportunities, risks and mitigations".* *https://www.ibm.com/downloads/documents/us-en/10a99803d8afd656*

> *because of the complex way in which the many variables in AI/machine learning models interact, the results of these models are often less interpretable than traditional statistical models.*

— *PNC Financial Services, PNC Financial Services 2025 Annual Report.* *https://d1io3yog0oux5.cloudfront.net/_dffbb0389228153a0fb81378bd61001a/pnc/db/2452/24001/annual_report/PNC+2025+Annual+Report_ADA.pdf*

## 7.5 AI welfare and rights

Risk definition: Ethical considerations regarding the treatment of potentially sentient AI entities, including discussions around their potential rights and welfare, particularly as AI systems become more advanced and autonomous.

No company in the sample was coded as acknowledging this risk, so there are no excerpts.

## 7.6 Multi-agent risks

Risk definition: Risks from multi-agent interactions, due to incentives (which can lead to conflict or collusion) and/or the structure of multi-agent systems, which can create cascading failures, selection pressures, new security vulnerabilities, and a lack of shared information and trust.

> *Also, background execution in multi-agent environments increases the risk of covert channels if input/output validation is weak.*

— *IBM, "AI agents: Opportunities, risks, and mitigations — IBM AI Ethics Board".* *https://www.ibm.com/granite/docs/resources/ai-agents-opportunities-risks-and-mitigations.pdf*

# Exploratory mitigations analysis

We conducted an exploratory analysis to identify and classify AI risk mitigations publicly reported by the largest global companies. A mitigation report was defined as a statement in which the company describes an action that reduces the likelihood or impact of an AI risk.

We initially intended to analyze and describe mitigation reports in the main text, by organizing mitigations into categories using a taxonomy, and linking the risks companies acknowledged with the mitigations they reported addressing those risks. However, the unreleased draft taxonomy we used in analysis was significantly revised after classification. In addition, almost all mitigations reported by companies were not reported in sufficient detail to determine the relevant risk addressed.

As with risk acknowledgements, passages from company documents that mentioned AI and/or AI risk were classified by an LLM (GPT-5.2) using a structured prompt with two stages: first, whether any mitigation was reported; second, what specific mitigation(s) were reported. Passages describing AI products or capabilities without mentioning mitigations, generic statements about mitigations not explicitly linked to AI (e.g., "cybersecurity"), and material reflecting the positions of external parties (such as shareholder proposals) were not classified as mitigation reports.

We validated both stages of the model's classifications against independent human coding using the same stratified random sample of 500 passages mentioned in the Methods section. Agreement on whether any mitigation was reported was adequate (model-human $\kappa$ 0.64-0.87; human-human 0.56-0.84). However, agreement on the specific mitigation was weak (model-human 0.25-0.48; human-human 0.24-0.42). We believe this is due to the draft status of the taxonomy making the mitigation categories difficult to distinguish, as well as the lack of information in mitigation reports to support classification at this level. For this reason, we only report the summary finding of whether companies reported *any* AI risk mitigations.

Of the largest 100 companies worldwide, 71% reported one or more AI risk mitigation, and we estimate that 38.6% (95% CI [30.7%, 46.5%]) of the largest 1000 companies did so.